\documentclass[aps,prx,amsmath,amssymb,twocolumn,superscriptaddress,floatfix,10pt]{revtex4-2}
\pdfoutput=1

\usepackage{mathtools}
\usepackage[english]{babel}

\usepackage{graphicx}
\usepackage{bm}
\usepackage{siunitx}

\usepackage{lipsum}
\usepackage{complexity}

\usepackage{glossaries}
\glsdisablehyper
\newacronym{mps}{MPS}{matrix product states}
\newacronym{mpo}{MPO}{matrix product operator}
\newacronym{ed}{ED}{exact diagonalization}
\newacronym{tns}{TNS}{tensor network states}
\newacronym{eop}{EOP}{entanglement of purification}
\newacronym{tdvp}{TDVP}{time-dependent variational principle}
\newacronym{tebd}{TEBD}{time-evolving block decimation}
\newacronym{metts}{METTS}{minimally entangled typical thermal states}
\newacronym{itp}{ITP}{imaginary time propagation}

\usepackage{algorithm}
\usepackage{algpseudocode}

\usepackage{hyperref}
\usepackage[capitalize]{cleveref}

\DeclareMathOperator{\trace}{tr}

\newcommand\bra[1]{\langle #1 \vert}
\newcommand\ket[1]{\vert #1 \rangle}
\newcommand\braket[2]{\langle #1 \vert #2 \rangle}
\newcommand\ketbra[1]{\ket{#1}\bra{#1}}
\newcommand\set[1]{\{\,#1\,\}}
\newcommand\norm[1]{\|#1\|}
\newcommand\abs[1]{|#1|}
\newcommand\order[1]{\mathcal{O}(#1)}

\newcommand\Pauli{\mathcal{P}}

\renewcommand\vec\boldsymbol

\newcommand\threesat{3-SAT}
\newcommand\sharpthreesat{$\sharp$\threesat}

\newcommand\package[1]{\textsc{#1}} 

\usepackage{orcidlink}

\begin{document}

\title{Entanglement Barriers from Computational Complexity: Matrix-Product-State Approach to Satisfiability}

\newcommand\tudresdenaffil{Institute of Theoretical Physics, Technische Universit\"at Dresden, 01062 Dresden, Germany}
\newcommand\pksaffil{Max Planck Institute for the Physics of Complex Systems, N\"{o}thnitzer Str. 38, 01187 Dresden, Germany}
\newcommand\tumaffil{Technical University of Munich, TUM School of Natural Sciences, Physics Department, 85748 Garching, Germany}
\newcommand\mcqstaffil{Munich Center for Quantum Science and Technology (MCQST), Schellingstr. 4, 80799 M\"unchen, Germany}
\author{Tim Pokart\,\orcidlink{0009-0006-1415-3261}}
\affiliation{\tudresdenaffil}
\author{Frank Pollmann\,\orcidlink{0000-0003-0320-9304}}
\affiliation{\tumaffil}
\affiliation{\mcqstaffil}
\author{Jan Carl Budich\,\orcidlink{0000-0002-9859-9626}}
\affiliation{\tudresdenaffil}
\affiliation{\pksaffil}

\begin{abstract}
    We approach the \threesat{} satisfiability problem with the quantum-inspired method of \gls{itp} applied to \gls{mps} on a classical computer.
    This ansatz is fundamentally limited by a quantum entanglement barrier that emerges in imaginary time, reflecting the exponential hardness expected for this $\NP$-complete problem.
    Strikingly, we argue based on careful analysis of the structure imprinted onto the \gls{mps} by the \threesat{} instances that this barrier arises from classical computational complexity.
    To reveal this connection, we elucidate with stochastic models the specific relationship between the classical hardness of the $\sharp \P \supseteq \NP$-complete counting problem \sharpthreesat{} and the entanglement properties of the quantum state.
    Our findings illuminate the limitations of this quantum-inspired approach and demonstrate how purely classical computational complexity can manifest in quantum entanglement. 
    Furthermore, we present estimates of the non-stabilizerness required by the protocol, finding a similar resource barrier.
    Specifically, the necessary amount of non-Clifford operations scales superlinearly in system size, thus implying extensive resource requirements of \gls{itp} on different architectures such as Clifford circuits or gate-based quantum computers.
\end{abstract}

\maketitle

\glsresetall
\section{Introduction}
Hard unresolved problems such as $\NP$-complete satisfiability problems and their quantum counterparts known as $\QMA$-complete problems are among the fundamental bottlenecks to the development of human knowledge~\cite{Aaronson_2013,aaronson_6555741}. 
While the advent of quantum computing promises decisive progress on certain classically unresolved problems, e.g., with Shor's algorithm making prime factoring computationally viable on a quantum computer~\cite{Shor94}, solutions to the aforementioned generic $\NP$- and $\QMA$-complete problems remain elusive. 
Yet, posing such problems as concrete quantum information processing tasks---for example by encoding their solution into the ground state $\ket{\psi_s}$ of a local problem Hamiltonian $H$---is well established, which rephrases the challenge into understanding (and where possible overcoming) limitations to quantum state preparation~\cite{hu2025asymptoticexceptionalsteadystates}. 
This issue may be approached from a purely quantum perspective envisioning quantum hardware~\cite{RevModPhys.94.015004,RevModPhys.90.015002,ADIABATICQUANTUMCOMPUTING2}, in a hybrid fashion with quantum machines assisted by classical computers~\cite{TILLY20221,farhi20A14quantumapproximateoptimizationalgorithm}, or in the framework of quantum-inspired algorithms using classical hardware to emulate quantum concepts such as coherent superposition and entanglement \cite{PhysRevX.14.041029,PRXQuantum.5.010308,pednault2019leveragingsecondarystoragesimulate,PhysRevX.10.041038}.

Interestingly, even a dynamical path to the problem solving ground state $\lvert \psi_s\rangle$ of $H$ may be generically identified in the framework of \gls{itp} as
\begin{align}
    \lvert \psi_s\rangle = \lim_{\tau \rightarrow \infty}\lvert \psi(\tau) \rangle \quad \mathrm{with} \quad \ket{\psi(\tau)} = \frac{\mathrm{e}^{-\tau H}\,\ket{\psi_0}}{\norm{\mathrm{e}^{-\tau H}\,\ket{\psi_0}}}, 
    \label{eq:imagtimeevolution}
\end{align}
where the initial state $\ket{\psi_0}$ is guaranteed to have non-zero overlap with $\ket{\psi_s}$, i.e., $\lvert\braket{\psi_s}{\psi_0}\rvert > 0$.
If the spectral gap $\Delta$ above the ground state manifold scales at least inverse polynomially in system size $n$, imaginary-time convergence occurs on polynomial timescales $\sim n/\Delta$. 
The prominent $\QMA$-complete \emph{local Hamiltonian problem} features polynomial gaps~\cite{doi:10.1137/S0097539704445226,Aharonov2009,watrous2008quantumcomputationalcomplexity}, while $\NP$-complete satisfiability problems can even be encoded in Hamiltonians with constant gap.
These observations allow us to naturally reformulate the crucial challenge into the question how implementing \gls{itp} is stymied by fundamental obstructions, both on quantum and classical devices. 
On the quantum side, \gls{itp} is well known to amount to expensive unitarization~\cite{Motta2020,McArdle2019,anglescastillo2025understandingquantumimaginarytime} or post-selecting dissipative processes in which no quantum jumps occur, which generically attenuates their statistical weight exponentially in time, thus requiring exponentially many experimental runs~\cite{chowdhury2016quantumalgorithmsgibbssampling}. 
On the classical side, stochastic (Monte Carlo) algorithms \cite{10.5555/3157382.3157634,PARALLELTEMPERING1,PARALLELTEMPERING2,SIMULATEDANNEALING,huynh2023quantuminspiredmachinelearningsurvey,Alvarez-Alvarado2021} are known to encounter exponentially long equilibration times that reestablish the exponential wall of computational complexity.

\begin{figure}
    \centering
    \includegraphics{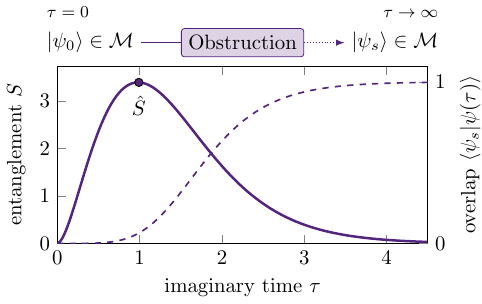}
    \caption{An initial state $\ket{\psi_0}$ is transformed into $\ket{\psi_s}$ containing the solution to a hard combinatorial problem by evolution of the imaginary time $\tau$. 
    Both states are unentangled and thus trivially representable in the MPS ansatz space $\mathcal{M}$. 
    The protocol is obstructed by a bump 
    of the half-chain entanglement entropy $S$ at intermediate imaginary times.
    The plot shows data for a uniquely solvable \threesat{} instance with $n=18$ variables and $m=76$ clauses ($\alpha=4.22$). 
    Solving the combinatorial problem as measured by the total weight of the solution (dashed) is only possible after facing the entanglement bump of height $\hat{S}$ at time $\hat{\tau}$.}
    \label{fig:illu-entanglement-bump}
\end{figure}

Here, we employ deterministic variational wave-function methods known as \gls{mps}~\cite{Fannes1992,SCHOLLWOCK201196} which approximately encode the imaginary time evolved quantum state $\ket{\psi(\tau)}$ in \cref{eq:imagtimeevolution}. 
They are particularly appealing for solving combinatorial problems, where both an unbiased candidate for the initial state $\ket{\psi_0}$ and the solution states $\ket{\psi_s}$ are readily contained in the variational \gls{mps} manifold.
For simulating real time evolution, \gls{mps} are hindered by temporal entanglement barriers~\cite{PhysRevResearch.6.043077,PhysRevB.105.075131,PhysRevB.110.245109}. 
We demonstrate a similar phenomenon, in which an entanglement bump in imaginary time $\tau$ of extensive height culminating in $\hat{S}(\hat{\tau})$ (see Fig.~\ref{fig:illu-entanglement-bump}) generically needs to be overcome before the solution to the problem can be determined from $\ket{\psi(\tau)}$. 
A priori this does not forbid the applicability to \gls{itp} for combinatorial problems, as any approximation errors incurred from the restrictions of the ansatz space may be remedied by the exponential amplification of the solutions amplitude. 
To elucidate suppression of this apparent saving grace, we first link key characteristics describing the computational complexity of the satisfiability problem's counting variant \sharpthreesat{}, such as the critical density $\alpha$ of combinatorial constraints (clause density) at which the problem is hardest, directly to the aforementioned entanglement barrier. 
This allows us to study in depth how the MPS approximation to $\ket{\psi(\tau)}$ for the $\NP$-complete satisfiability problem \threesat{} still eludes convergence to the solution.
We both find a statistical model to describe the entanglement and thus informational content encoded in the \gls{mps} and provide a tractable explanation of the hardness transition encountered by the \gls{mps} thus complementing previous mathematical approaches studying  the complexity of \sharpthreesat{}~\cite{BAILEY20071627} unrelated to quantum entanglement.
In this context, we also discuss whether the \gls{mps} ansatz can provide an optimally compressed proof certificate, thus book-keeping the complete list of solutions to a \sharpthreesat{}  instance.

The remainder of this article is structured as follows: in \cref{sec:1}, we outline the imaginary time evolution approach to solve classical combinatorial problems and describe the entanglement barrier prohibiting its practical application.
In \cref{sec:2}, we focus on a simplified imaginary time evolution to explain how this entanglement can be seen as an artifact of the computational complexity encoded in the \gls{mps} and work out constraints put on the quantum inspired variational approach by the combinatorial problem.
We conclude the paper in \cref{sec:conclusion}.
We used \package{TeNPy}~\cite{tenpy2024}, \package{ITensors.jl}~\cite{10.21468/SciPostPhysCodeb.4,10.21468/SciPostPhysCodeb.4-r0.3}, \package{KrylovKit.jl}~\cite{haegeman_2025_17088820}, \package{Optim.jl}~\cite{mogensen2018optim}, and \package{PicoSAT}~\cite{doi:10.3233/SAT190039} for computations.

\section{Imaginary Time Propagation}\label{sec:1}

Our main focus is on the boolean satisfiability problem \threesat.
It is of theoretical interest to study the class of $\NP$-complete problems~\cite{10.1145/800157.805047,Karp1972} and has practical applications, e.g., in bringing reasoning to artificial intelligence~\cite{pmlr-v97-wang19e}, in computational biology~\cite{10.1007/978-3-030-42266-0_6} and as proof framework in mathematics~\cite{10.1007/978-3-319-40970-2_15}.

An instance of \threesat{} is defined for $n$ binary variables $x_1, \ldots x_n$ constrained by a set of $m$ clauses $C_1, \ldots, C_m$.
Each clause contains exactly three distinct variables (which may be negated) joined together by the logical disjunction ($\lor$), such that assigning one of the variables correctly renders the whole clause satisfied.
The clauses are then joined together by the logical conjunction ($\land$) to obtain the conjunctive normal form (CNF) $\Phi = C_1 \land \ldots \land C_m$; satisfying the CNF $\Phi$ is the objective of the problem.
The clause to variable ratio $\alpha = m / n$ describes the hardness of the satisfiability problem, at which the CNF instances turn from typically satisfiable to typically unsatisfiable.
For \threesat{} this transition occurs at $\alpha_c \approx 4.27$~\cite{SATUNSATTRANSITION1,SATUNSATTRANSITION2,REPLICAMETHOD1}.
All instances used in this manuscript are generated to be unbiased but satisfiable~\cite[$p_0=0.08$]{doi:https://doi.org/10.1002/3527603794.ch7,PhysRevLett.88.188701}. 

This intrinsically classical problem can be lifted to the realm of quantum computation by replacing each binary variable with a spin variable $x_i \rightarrow \sigma_i^z$~\cite{PhysRevX.14.041029,PhysRevA.74.060304}, thus associating logical true (false) with spin up (down).
By finding a formulation of the search function in terms of a Hamiltonian $H$, obtaining the answer to the satisfiability problem then corresponds to preparing the ground state of $H$.
In the case of \threesat{} such a mapping can be constructed explicitly by considering for each clause $C_j = x_{j_1} \lor \ldots \lor x_{j_3}$ an encoding in terms of a $3$-local Hamiltonian $h_j$ as
\begin{align*}
    h_j = \prod_{i=1}^3 \frac{1 - \mathrm{sign}(x_{j_i}) \sigma^z_{j_i}}{2}.
\end{align*}
For any quantum state $\ket{x} = \ket{x_1\ldots x_n}$, we have
\begin{align*}
    h_j \ket{x} = \begin{cases}
                      0,       & \text{if } x \text{ satisfies } C_j \\
                      \ket{x}, & \text{if } x \text{ violates } C_j.
                  \end{cases}
\end{align*}
Then $H = \sum_j h_j$ assigns to each computational basis state $\ket{x}$ the energy $E_\Phi(x)$ given by the total number of clauses violated in $\Phi$ by $x$.

A search for the satisfying assignment can now be conducted, e.g., by performing Grover search~\cite{GROVERSEARCH,PhysRevX.14.041029}, a quantum approximate optimization~\cite{PhysRevA.61.052311,PRXQuantum.5.030348}, quantum Monte Carlo~\cite{PhysRevA.83.012309,PhysRevApplied.17.024052,PhysRevApplied.23.034031}, using adiabatic quantum computing~\cite{ADIABATICQUANTUMCOMPUTING1,ADIABATICQUANTUMCOMPUTING2,ADIABATICQUANTUMCOMPUTING3}, Zeno dragging~\cite{zhang2025optimalschedulemultichannelquantum}, quantum annealing~\cite{10.1007/978-3-030-14082-3_3} or by leveraging quantum inspired heuristics~\cite{preisser2025variationalmatrixproductstates}.
In this paper, we extend this collection by studying an algorithm to solve the combinatorial problem through imaginary time evolution.
Starting from the product state $\ket{\psi_0} = \ket{+_1\ldots +_n}$ initially polarized perpendicular to the computational basis and following the time evolution in \cref{eq:imagtimeevolution} yields the solution subspace in linear imaginary time $\tau = \order{n}$.
This approach is however limited by an intermediate entanglement bump as illustrated in \cref{fig:illu-entanglement-bump}: while both initial and final state are product states, at some intermediate imaginary time $\hat{\tau}$ at which the solution can not yet be extracted, the entanglement reaches its maximum.

\begin{figure}
    \includegraphics{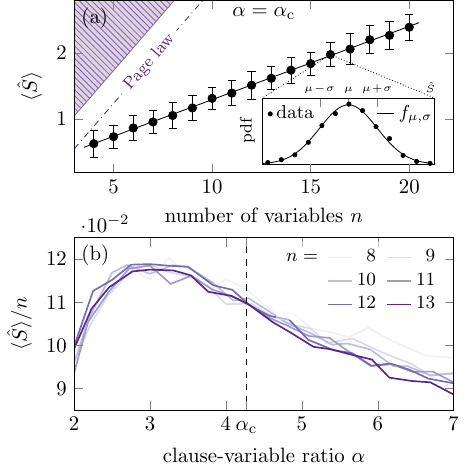}
    \caption{Maximal height $\hat{S}$ of the half-chain entanglement bump as illustrated in \cref{fig:illu-entanglement-bump} averaged over $1000$ instances. In (a) the scaling of the mean $\mu = \langle \hat S \rangle$ with the number of variables is shown to be linear at $\alpha=\alpha_c \approx 4.27$ while the standard deviation $\sigma$ (error bars) remains constant. Compared to the maximally possible entropy $S_{\mathrm{max}} = n \ln 2 / 2$ (hatched region) and that of a typical random states given by the Page law $S_{\mathrm{typ}} \approx S_{\mathrm{max}} - 1/2$~\cite{PhysRevLett.71.1291} (dash-dot line), the entanglement bump height is smaller as discussed in \cref{sec:stat_entanglement_model}. The inset confirms that $\mu$ and $\sigma$ are sufficient to characterize the distribution of $\hat S$ by comparing the numerically obtained probability density function to a Gaussian $f_{\mu,\sigma}$ ($\mu=\num{2.181}, \sigma=\num{0.207}$).  In (b) the mean of $\hat{S} / n$ is shown for a range of interesting $\alpha$ values stressing the universal properties independent of system size. Note that $S_{\mathrm{max}} / n = \ln 2 / 2 \approx \num{0.15}$. The temporal position of the entanglement bump $\hat{\tau}$ is investigated in \cref{app:entanglement_scaling} .}
    \label{fig:properties-entanglement-bump}
\end{figure}

In \cref{fig:properties-entanglement-bump} we measure the height $\hat{S}$ of the entanglement bump for various system sizes, finding $\hat{S}$ to scale linear with system size while its functional dependence on $\alpha$ converges towards a common form.
This is in agreement with the $\NP$-complete nature of the problem: the exponential time conjecture~\cite{EXPTIMECONJECTURE} implies that the time complexity of imaginary time evolution on a classical computer should scale as $\order{2^{cn}}$ with some constant $c$.
The complexity of simulating the imaginary time evolution on a classical computer is proportional to the entanglement entropy of the state~\cite{PhysRevLett.100.030504}.
Thus, we expect $S$ to grow linearly in system size, i.e., the entanglement is extensive in the number of variables $n$.
As we expect problems with similar $\alpha$ to be similarly hard, the uniform scaling in $S(\alpha, n) =  n f(\alpha)$ is also expected.
While the entanglement entropy continues to grow for $\alpha < \alpha_c$, in this region information about a satisfying state can be extracted through measurement of the state much earlier.
The entanglement bump itself starts to shrink only for much smaller $\alpha$; hence, the complexity of solving the \threesat{} problem is not encoded in the entanglement entropy itself, which however reflects properties of a different computational problem as described in \cref{sec:3}.
Lastly, we note an important difference of the quantum inspired approach when compared to classical algorithms, where its often meaningful to distinguish typical instances from worst-case ones. 
The former is thought to often contain enough structure, such that industrial grade instances are accessible using specialized solvers; see Ref.~\cite{donald2015art} for satisfiability problems, and for more general discussions Refs.~\cite{johnson1979computers,Cormen2022-xl}. 
In the quantum case, the nature of the entanglement bump becomes almost instance independent already for small system sizes $n \simeq 20$ with the standard deviation of $\hat S$ vanishing compared to its mean, cf.~the inset in \cref{fig:properties-entanglement-bump}\,(a). 
This suggests that the quantum inspired approach shows only weak instance dependence instead of the strong instance dependence in the classical realm.
The critical time $\tau^\star$ at which the entanglement bump occurs is independent of system size as shown in \cref{app:entanglement_scaling}.

\begin{figure}
    \centering
    \includegraphics{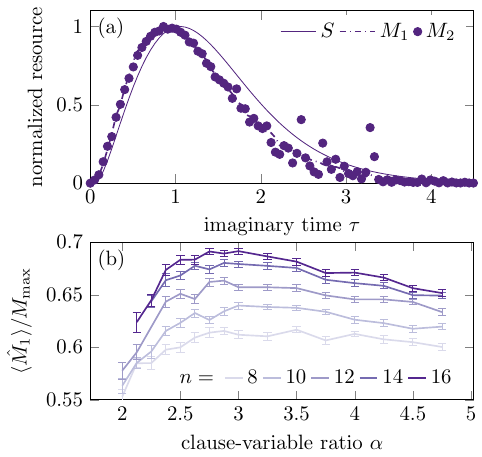}
    \caption{In (a) the entanglement entropy $S$ and the $\alpha=1$ ($\alpha=2$) stabilizer Rényi entropy $M_1$ ($M_2$) for a uniquely solvable \threesat{} instance with $n=15$ variables at $\alpha_c$; for the stabilizer Rényi entropies $\num{1000}$ ($\num{200000}$) samples were drawn to estimate them. Both measures of non-stabilizerness feature the same characteristics as the entanglement entropy. In (b) we show the average non-stabilizerness $\langle \hat M_1\rangle$---normalized to its maximal value $M_{\mathrm{max}} = n \ln 2$---and its uncertainty at different clause to variable ratios $\alpha$ sampled from $100$ instances for different number of variables $n$. Lines are provided for visual guidance only.}
    \label{fig:magic}
\end{figure}

While entanglement is an adequate indicator of \gls{mps} representability, highly entangled states are not sufficient indicators of quantum states hard to simulate on classical computers~\cite{gottesman1998heisenbergrepresentationquantumcomputers,PhysRevA.70.052328,PhysRevA.71.022316}. 
Recently, measures of the non-stabilizerness of quantum states have established themselves as alternative means to quantify the complexity of quantum states~\cite{PhysRevB.110.045101}; this complexity measure is fundamentally different from entanglement as it is a global property of the state independent of bipartitions.
Here, we provide measurements of the $\alpha$ stabilizer Rényi entropy~\cite{PhysRevLett.128.050402}
\begin{align*}
    M_\alpha(\ket{\psi}) \equiv \frac{1}{1 - \alpha} \ln \Xi_\alpha(\ket{\psi}),
\end{align*}
where $\Xi_\alpha(\ket{\psi})\equiv \sum_{P\in \Pauli_n} \abs{\braket{\psi|P|\psi}}^{2\alpha} / 2^n$ is the stabilizer purity computed from Pauli string $P$ in the Pauli group $\mathcal{P}_n$.
The Rényi entropies are bounded from above as $M_\alpha < M_{\mathrm{max}} = n\ln 2$~\cite{PhysRevLett.128.050402}.
For $\alpha=1$ the stabilizer entropy $M_1$ may be computed using perfect Pauli sampling~\cite{Haug2023stabilizerentropies,PhysRevLett.131.180401}, while for $\alpha=2$ we employ Pauli-Markov \gls{mps}~\cite{PRXQuantum.4.040317}.
Note that while $M_2$ is a magic monotone~\cite{Obst_2024,Leone_2024}, $M_1$ is not~\cite{Haug2023stabilizerentropies}.
In \cref{fig:magic}\,(a), we compare the normalized measures of non-stabilizerness, i.e., the quantities scaled such that their maximum is unity, to the entanglement entropy.
There, we again find a non-stabilizerness bump in imaginary time $\tau$, similar to the entanglement bump observed in \cref{fig:illu-entanglement-bump}.
While the height of the entanglement bump grows linear in system size, the scaling of the maximal non-stabilizerness $\langle \hat M_1\rangle$ measured during the imaginary time evolution suggests superlinear convergence towards its maximal value $M_{\mathrm{max}}$, cf.~\cref{fig:magic}\,(b).
As the stabilizer Rényi entropies provide a lower bound on the number of $T$-gates necessary to implement the unitary circuit synthesizing the intermediate states of \gls{itp}~\cite{PhysRevLett.128.050402,PhysRevApplied.19.034052}, this implies extensive resource requirements of such approaches.
Consequently, the simulability of \gls{itp} on contemporary gate-based quantum computers is constrained as their runtimes are dominated by these non-Clifford gates~\cite{gidney2024magicstatecultivationgrowing,Ruiz2025,PRXQuantum.4.010301}.
The functional dependence of $\hat m_1$ on the clause-variable ratio $\alpha$ is qualitatively similar to that of the entanglement, with a peak at $\alpha\approx2.6$.
Therefore, both entanglement and non-stabilizerness observed are correlated~\cite{szombathy2025independentstabilizerrenyientropy} and capture the intrinsic hardness of quantum states obtained by the protocol in \cref{eq:imagtimeevolution}; in what follows we focus on entanglement as the principal indicator of hardness in the context of \gls{mps}.

\section{Quantum Entanglement from Computational Complexity}\label{sec:2}

While in the previous section the imaginary time evolution was performed continuously, the relationship between the quantum entanglement bump and the classical computational complexity of the problem manifests itself clearer in the limit $\tau \rightarrow \infty$. 
To obtain this limit, we adjust the protocol in the following way: instead of associating to each computational basis state $\ket{x}$ as energy $E_\Phi(x)$ the number of clauses $x$ violates in the problem instance $\Phi$, we construct a Hamiltonian $\tilde H$ with energies
\begin{align*}
    \tilde E_\Phi(x) = \begin{cases}
                           0, & \text{if } E_\Phi(x) = 0, \\
                           1, & \text{if } E_\Phi(x) > 0. \\
                       \end{cases}
\end{align*}
The flat spectrum variant can be realized by replacing the sum over clauses with a product thus constructing $\tilde H = 1 - \prod_j \tilde h_j$ with $\tilde h_j = 1 - h_j$.

This setup is similar to Grover search~\cite{10.1145/237814.237866} in that it only distinguishes solution and non-solutions, hence the entanglement bump bottleneck vanishes, as the resulting state only has at most $\ln 2$ entanglement.
However, the problem shifts toward finding an applicable representation of $\tilde H$ as it is not obtained  as a computationally tractable sum of local terms anymore.
Naively implementing the action of $\tilde H$ by applying the product $\prod_j \tilde h_j$ term by term creates intermediate entanglement bumps of similar magnitude as those observed during the imaginary time evolution in \cref{sec:1}~\cite{PhysRevX.14.041029}.
In the remainder of this section we will investigate the nature of this intermediate entanglement bump, intricately related to the fact that one does not merely solve the $\NP$-complete \threesat{} problem, but instead the $\PP$-complete counting problem generalization of \threesat~\cite{garciasaez2011exacttensornetwork3sat,10.21468/SciPostPhys.7.5.060}.
    
\subsection{Contents of the wavefunction}

\begin{figure*}[t]
    \centering
    \includegraphics{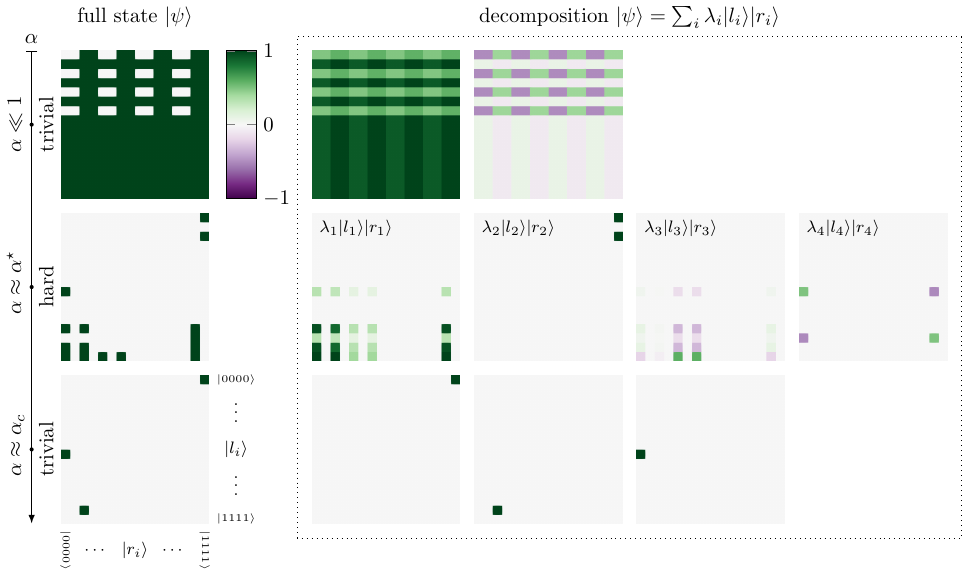}
    \caption{The three complexity regimes reflected in the structure of a typical \threesat{} quantum state $\ket{\psi}$ and its Schmidt decomposition for $n=8$ variables at different $\alpha$ ratios.
    The weights of $\ket{\psi}$ are shown with the computational basis spatially separated into left $\ket{l_i}$ (vertical) and right $\ket{r_i}$ (horizontal) states, which is the decomposition the \gls{mps} encodes at the central bond.
    For small clause-to-variable ratios $\alpha \ll 1$ (top row), the Schmidt states merely subtract weight for forbidden bitstrings from to the initial state, thus being similar to listing all states excluded from satisfying an instance. 
    Similarly, for ratios $\alpha \approx \alpha_c$ (bottom row) in which only few disjoint bitstrings remain viable, the Schmidt weights are formed by superpositions of their computational basis states.
    In the hard regime $\alpha \approx \alpha^\star$ (middle row), the Schmidt weights are formed by grouping residual correlations between the variables, leading to the formation of logically correlated states. In the example, the first, third and fourth Schmidt vectors group to represent the allowed computational basis states in the bottom half of the full state.}
    \label{fig:classical-approximation-illustration}
\end{figure*}

First, we want to elucidate the informational content of the wavefunction encoded by the \gls{mps}.
By successively applying the projectors corresponding to the different clauses, the final quantum state is
\begin{align*}
    \ket{\psi} = \sum_x \tilde E_\Phi(x) \ket{x} \equiv \sum_x \psi_{x_1\ldots x_n} \ket{x}.
\end{align*}
Here, the amplitudes $\psi_{x_1\ldots x_n}$ are unity if the bit string $x_1 \ldots x_n$ is a satisfying assignment to the CNF $\Phi$.
A different viewpoint on the amplitudes $\psi_{x_1\ldots x_n}$ is that of the full decision tree of the problem, where starting at $x_1$ each node we follow the left (right) edge if the variable is true (false).
Exactly this full decision tree is optimally represented by the \gls{mps}, i.e.,
\begin{center}
    \begin{tikzpicture}
        \begin{scope}[yshift=0.75cm,xshift=-2cm]
            \node[draw, circle, minimum width=0.2cm,inner sep=0,outer sep=0] (A) at (0,0) {};
            \node at (0,-0.35) {$x_1$};
            \node[draw, circle, minimum width=0.2cm,inner sep=0,outer sep=0] (B) at (-0.5,-0.5) {};
            \node[draw, circle, minimum width=0.2cm,inner sep=0,outer sep=0] (C) at (+0.5,-0.5) {};

            \draw[-latex] (A) -- (B) node[above left, midway] {0};
            \draw[-latex] (A) -- (C) node[above right, midway] {1};

            \node[draw, circle, minimum width=0.2cm,inner sep=0,outer sep=0] (D) at (-1.0,-1.0) {};
            \node at (-1.0,-1.35) {$x_n$};
            \node[draw, circle, minimum width=0.2cm,inner sep=0,outer sep=0] (E) at (-1.5,-1.5) {};
            \node[below=0.1cm] at (-1.5,-1.5) {$\psi_{0\ldots 0}$};
            \node[draw, circle, minimum width=0.2cm,inner sep=0,outer sep=0] (F) at (-0.5,-1.5) {};
            \node[below=0.1cm] at (-0.5,-1.5) {$\psi_{0\ldots 1}$};

            \draw[-latex, densely dotted] (B) -- (D);

            \draw[-latex] (D) -- (E) node[above left, midway] {0};
            \draw[-latex] (D) -- (F) node[above right, midway] {1};
        \end{scope}

        \node at (0,0) {$\xleftrightharpoons[\text{is data}]{\substack{\text{is optimal}\\\text{representation}}}$};

        \begin{scope}[xshift=1.5cm,yshift=-0.25cm]
            \draw[rounded corners] (0.25,0.25) circle (0.25);
            \draw[rounded corners] (1.00,0.25) circle (0.25);
            \draw[rounded corners] (1.75,0.25) circle (0.25);
            \draw[rounded corners] (2.50,0.25) circle (0.25);

            \draw (0.50,0.25) -- ++(0.25,0);
            \draw[densely dotted] (1.25,0.25) -- ++(0.25,0);
            \draw (2.00,0.25) -- ++(0.25,0);

            \draw (0.25,0) -- ++(0,-0.25) node[below] {$x_1$};
            \draw (1.00,0) -- ++(0,-0.25);
            \draw (1.75,0) -- ++(0,-0.25);
            \draw (2.50,0) -- ++(0,-0.25) node[below] {$x_n$};
        \end{scope}
    \end{tikzpicture}
\end{center}
Here, optimal is to be understood in the following way: at each bond, the \gls{mps} finds the lowest rank approximation achievable for the joint probability distribution of bitstrings to its left and right which still yield a satisfying assignment.
Since the decompositions involved in \gls{mps} require floating point numbers, the interpretation of the basis states to the left (or right) in terms of decision trees is not straightforward, as their values are real and can even be negative.
Hence, once the \gls{mps} gets truncated the aforementioned compressed decision tree picture gets fuzzy.
In \cref{app:boolean_compression}, we compare this floating point truncation to a Boolean truncation formalism. 

The above construction highlights the shortcomings of quantum inspired approaches seeking a faithful representation of the wavefunction to solve combinatorial problems.
Computing the normalization $\norm{\ket{\psi}}$ immediately yields the number of computational basis states the wavefunction includes.
In this way, one does not merely provide any solution but instead answers the question of how many solutions still exist, incidentally solving the counting problem generalization of the satisfiability problem.
These $\sharp$SAT problems are $\sharp \P$ complete and hence much higher in the complexity hierarchy~\cite{watrous2008quantumcomputationalcomplexity}.
The related decision problem of deciding if at least $\sqrt{2^n}$ variable assignments satisfy a given CNF $\Phi$ is $\PP$-complete~\cite{BAILEY20071627}.
Hence, even some of the hardest problems for quantum computers belonging to $\QMA$ are subsumed by the problem the MPS is trying to describe~\cite[Sec.~V.4]{watrous2008quantumcomputationalcomplexity}.
Deciding the $\PP$-complete decision variant of $\sharp$SAT has its hardness transition in the range between $0.923$ and $\alpha_\sharp\approx 2.595$~\cite{BAILEY20071627}.

Remarkably, this hardness can be seen in the structure of the Schmidt decomposition the \gls{mps} contains as illustrated in \cref{fig:classical-approximation-illustration}.
For small ratios $\alpha \ll 1$ the state only slightly varies from the initial state $\ket{\psi_0}$; the Schmidt vectors are akin to a classical exclusion list, where a priori every computational basis state is deemed satisfying except for few combinations ruled out by the enforced clauses. 
This structure is comfortably within the variational manifold of the \gls{mps}.
Similarly, for those $\alpha \approx \alpha_c$ for which the $\NP$-complete problem is the hardest, the \gls{mps} in its Schmidt vectors simply enumerates the few still satisfying assignments.
The hardest regime therefore is located around some $0<\alpha \ll \alpha_c$.
Indeed, we find that after passing an $\alpha^\star \approx 2.56 < \alpha_\sharp$ derived below, the Schmidt decomposition can no longer group correlated logical assignments together and thus its advantage for small $\alpha \ll 1$ vanishes.
At the same time, there are still an extensive amount of satisfying assignments, thus, between $\alpha^\star$ and $\alpha_\sharp$, the \gls{mps} encounters the hardest states with still macroscopically many possible logical assignments and no feasible compression available.
In \cref{fig:classical-approximation-illustration}, this regime is characterized by small groupings of computational basis states with residual correlations still allowed by the partially enforced constraints.

We continue in \cref{sec:stat_entanglement_model} by deriving a statistical model of the entanglement observed in the \gls{mps} both relating it to purely classical information and predicting the hardest point for larger system sizes than practically accessible.
In \cref{sec:transition_critical_ratio} the critical ratio $\alpha^\star$ at which compressibility of the wave function gets the hardest is derived.
Conclusions for the simulability of \gls{itp} using \gls{mps} are presented in \cref{sec:3}.

\subsection{Statistical entanglement model}\label{sec:stat_entanglement_model}

To model the entanglement observed upon application of the projector terms, we first describe entanglement properties of purely random combinatorial wavefunctions and then add the structure imposed by the \threesat{} problem.
Robust agreement between the derived model and sampled instances confirms this classical description of the entanglement, thus allowing to predict the hardest regime for \gls{mps} approaches.

\subsubsection{Random combinatorial states}
First, consider a purely random combinatorial state $\ket{\phi} = \sum_x \phi_x \ket{x}$ with $\phi_x \in \set{0, 1}$ chosen independently and randomly.
Let $F = \sum_x \phi_x$ be the total number of ones in the state and let $f = \ln_2 F / n$ be the filling fraction, i.e., how many of the $2^n$ possible amplitudes are unity.
As detailed in \cref{app:diagonal_model} the entanglement for such states are dominated by the diagonal entries of the reduced density matrix~\cite{PhysRevLett.124.200602} and as such can be modeled using a Poisson process, in which the average entanglement is 
\begin{align*}
    \langle S \rangle = \ln F + \mathbb{E}_{X \sim \mathrm{Poisson}(\lambda)}[-\ln (X + 1)]
\end{align*}
with $\lambda = 2^{(f - 0.5)n}$ the expected number of nonzero amplitudes contributing to each entry in the reduced density matrix.
This unwieldy form is bounded from above and below by
\begin{align*}
     fn - \ln (1 + 2^{(1/2 - f)n}) \leq \langle S \rangle \leq \begin{cases}
                               fn,  & \text{if } f \leq 1/2, \\
                               n/2, & \text{if } f > 1/2.
                           \end{cases}
\end{align*}
Both limits coincide in the thermodynamic limit $n\rightarrow\infty$, such that for an infinite dimensional system the entanglement entropy grows linearly up until filling $f=1/2$ and then stays constant for $f < 1$.
At full-filling, it again vanishes.

\subsubsection{\threesat{} states}

\begin{figure}
    \includegraphics{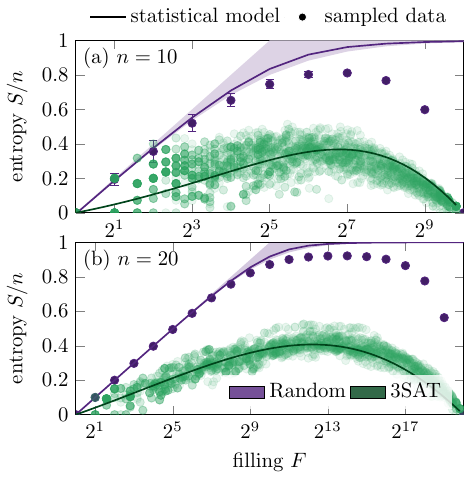}
    \caption{Entanglement entropy of the combinatorial random states (purple) and those states originating from constraints of \threesat{} (green). The data (dots) is shown along the respective statistical models (lines, bounds as area). States  corresponding to \threesat{} instances are differently entangled than random combinatorial ones. The random combinatorial state approaches the dominant diagonal model as described above in the thermodynamic limit; compare, e.g., $n=10$ and $n=20$. Similarly, the model for the \threesat{} entanglement in \cref{eq:markov_chain_model} strongly agrees with the numerical data.}
    \label{fig:entanglement-comparison-combinatorial-3sat}
\end{figure}

In \cref{fig:entanglement-comparison-combinatorial-3sat} we compare the entanglement observed in random combinatorial states and their stochastic model to \threesat{} instances.
The latter feature much lower entanglement, which is fully attributable to the classical correlations imposed on the wave function by the underlying combinatorial problem.
Accounting for this discrepancy involves two ingredients as described below: first, the partition of the total Hilbert space $\mathcal H$ into a left ($\mathcal L$) and right ($\mathcal R$) subsystem are thought of as continuously depleted entanglement reservoirs.
Initially, they are able to fully entangle the state, but as more combinations of variables are ruled out, their entanglement power is reduced.
Second, not all spins in the two subsystems have been entangled; at the start all spins are in product states and then progressively entanglement is built up by correlating the variables.
The two mechanisms act in contrary fashion: while applying more terms shrinks the entanglement reservoirs, it increases the correlations between the variables.
In the end, the total entanglement after applying $m$ constraints will be given by
\begin{align}
    S_m = C_m \ln \dim \mathcal L_m, \label{eq:markov_chain_model}
\end{align}
where $C_m$ is the amount of entangling correlations and $\ln \dim \mathcal L_m$ is the capacity of the (smaller) entanglement reservoir.

\paragraph*{The entanglement reservoirs.}
Assuming that all clauses of a \threesat{} instance are generated independently, each projector (or constraint) removes on average $7/8$ of the possible combinations of variables.
Hence, the total unconstrained Hilbert space dimension behaves as
\begin{align}
    \dim \mathcal H_m = (7/8)^m \dim \mathcal H.
    \label{eq:reservoir_dimension}
\end{align}
In the case of a bipartition into two even and equal subsystems, we find that
\begin{align*}
    \dim \mathcal L_m = \dim \mathcal R_m = \sqrt{\dim \mathcal H_m} = \left(7/8\right)^{m/2} \sqrt{2^n}.
\end{align*}
Here, we associated the Hilbert space dimension directly to the number of remaining satisfying assigments and thus the filling $f$ of the state; thus an intricate relation between the critical value $\alpha$ and the filling of the state $f$ appears.
Relating $\dim \mathcal H_m = 2^{fn}$ and \cref{eq:reservoir_dimension} one obtains
\begin{align*}
    (f - 1) \ln 2 = \alpha \ln \frac{7}{8}.
\end{align*}
At filling $f_\sharp=1/2$ one recovers the $\PP$-transition upper bound $\alpha_\sharp$.

\paragraph*{The entangling correlations.}

\begin{figure}[t]
    \centering
    \includegraphics{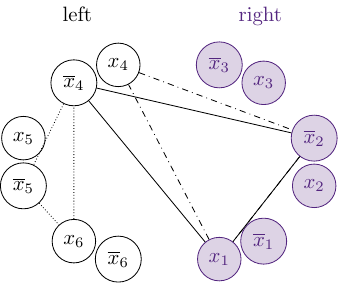}
    \caption{The entanglement graph for $6$ variables $x_1, \ldots, x_6$. The exclusion triangles for the clauses $x_1 \lor \overline x_2 \lor \overline x_4$ (solid), $x_1 \lor \overline x_2 \lor x_4$ (dash dot) and $\overline x_4 \lor \overline x_5 \lor x_6$ (dotted) are shown. While the first two are able to generate entanglement between the two subsystems, the third one is not. Furthermore, as the first two contain the same variable but with different polarities, their entanglement contribution should be ignored, as $(x_1 \lor \overline x_2 \lor \overline x_4) \land (x_1 \lor \overline x_2 \lor x_4) = x_1 \lor \overline x_2$, which is accounted for by the correction factor $f$.}
    \label{fig:graph}
\end{figure}

To model the entangling correlations, we consider the graph illustrated in \cref{fig:graph}.
There, each variable and its negation is assigned a node and clauses are included by drawing triangles connecting the three involved nodes.
Hence, the graph is full if every node is connected to every other except the one representing the same variable with opposite polarization.

The amount of entangling correlations is then modeled by the number of active edges across the cut representing the bipartition of the system.
Their number can be obtained using the following Markov chain approach: Let $N_d$ be the number of vertices of degree $d$, i.e., the number of nodes with $d$ outgoing edges.
Initially, no node has any outgoing edge, hence $N_0 = 2n$ and $N_{d>0} = 0$; any node may have at most $E = 2n-2$ outgoing edges.
When adding a constraining clause to the system, three nodes are selected at random and connected by a triangle.
For any selected node, it may gain two additional edges, one additional edge (if parts of the clause are common with a previous one) or no additional edge (if the clause has been drawn already).
For $d$ the degree of the node, there are $\tilde d = E - d$ possible additional edges that may be set.
We find the transition weights to gain zero $\Omega_0$, one $\Omega_1$ or two $\Omega_2$ additional edges as $\Omega_i = \max(0, \omega_i)$ with raw weights
\begin{align*}
    \omega_0 & = \tilde d (\tilde d - 1), &
    \omega_1 & = 2 \tilde d d,                     &
    \omega_2 & = d (d - 1).
\end{align*}
The transition probabilities are $p_i = \Omega_i / \sum_j \Omega_j$ and the selection probability of one node is $p_s = 3 / 2n$.
Then, for each additional constraint the node counts are updated as $N_{d+i}' \leftarrow p_s p_i N_d$ for $0 \leq i \leq 2$ from being selected and increasing the edge count and  $N_d' \leftarrow (1-p_s) N_d$ from not being selected.
In total, there will be $n_{\mathrm{active}} = \sum_j n_j j$ active edges in the graph, such that the current capacity can be computed as
\begin{align*}
    C = \frac{1}{2} \times \frac{n_{\mathrm{active}}}{n_{\mathrm{tot}}} \times \frac{n_{\mathrm{cut}}}{n_{\mathrm{tot}}} \times f(n_{\mathrm{active}}).
\end{align*}
The $1/2$ is to compensate the double counting of edges and $n_{\mathrm{tot}} = 2n (2n - 2) / 2$ is the total number of edges, while $n_{\mathrm{cut}} = n_{\mathrm{tot}} - 2 n (n - 2)$ is the number of edges that go over the cut bipartitioning the system.
Corrections from clauses canceling each others entangling effects as illustrated in \cref{fig:graph} are accounted for by the additional correction factor $f(n_{\mathrm{active}})$ as derived in \cref{app:correlation_correction}.
Comparing with \cref{fig:entanglement-comparison-combinatorial-3sat}, the entanglement predicted from this model matches that observed empirically.

\begin{figure}
    \includegraphics{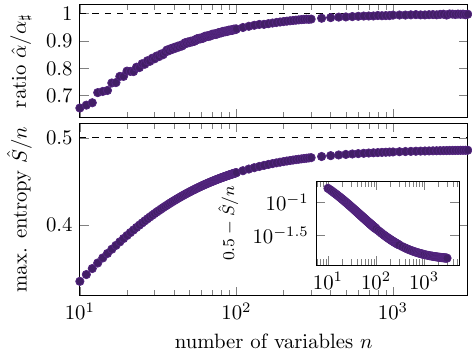}
    \caption{The critical clause-variable ratio $\hat{\alpha}$ and maximal entanglement $\hat{S}$ observed from the Markov chain model in \cref{eq:markov_chain_model}. The critical ratio $\hat{\alpha}$ approaches the upper limit of the classical hardness transition $\alpha_\sharp$, while the maximal entropy $\hat{S}$ approaches $n \ln_2(2)/ 2$ at which compression with \gls{mps} becomes unfeasible.}\label{fig:markov-model-critical-ratio-max-entropy}
\end{figure}

\subsubsection{Conclusions}

Based on the above model, we draw the following conclusions.
First, it is intriguing to use classical knowledge to assist the \gls{mps} approach, e.g., by finding an optimal variable to spin assignment.
However, finding a low entanglement cut through the graph in \cref{fig:graph} is related to MaxCut, more concretely to the Maximum Bisection Problem, which is known to be $\NP$-hard~\cite{GAREY1976237}.

Second, the \gls{mps} can be interpreted as a compressed representation of the full decision tree, such that it may be a viable alternative to an exhaustive list when trying to prove the answer to the $\PP$-problem.
By supplying a \gls{mps}, a verifier could (i) check that applying the projectors corresponding to the clause does not alter the \gls{mps} and (ii) compute the norm of the \gls{mps} to verify the total number of solutions.
\Cref{fig:markov-model-critical-ratio-max-entropy} depicts the maximal entanglement observed in the model in \cref{eq:markov_chain_model}.
We expect the \gls{mps} bond dimension to scale as $\chi \propto 2^{\hat{S}}$ and with the number bits required to store the representation scaling as $\propto \chi^2$.
Since the entanglement saturates close to  $\hat{S} / n \approx 0.5$, the \gls{mps} does not provide a significant compression gain compared to an exhaustive list of solutions for large $n$.

Third, by finding the critical ratio $\hat{\alpha}$ at which the entanglement entropy peaks, we find the hardest point at which to obtain the \gls{mps} representation.
This critical ratio $\hat{\alpha}$ for large $n$ approaches the upper limit $\alpha_\sharp$ of the possible range at which the classical problem becomes the hardest, cf.~\cref{fig:markov-model-critical-ratio-max-entropy}.

\subsection{On the Transition at the Critical Ratio}\label{sec:transition_critical_ratio}

With the structures generating the entanglement better understood, the qualitative difference between states at both sides of the complexity transition remains illusive.
For this, consider again the illustration of a typical quantum state and its Schmidt decomposition after the complexity transition in \cref{fig:classical-approximation-illustration}.
Near the hardest point the states are sparsely populated; nevertheless some Schmidt states  group highly correlated satisfying configurations.
Exactly this grouping becomes almost surely impossible after surpassing some critical ratio $\alpha^\star$ as derived below.

\begin{figure}
    \includegraphics{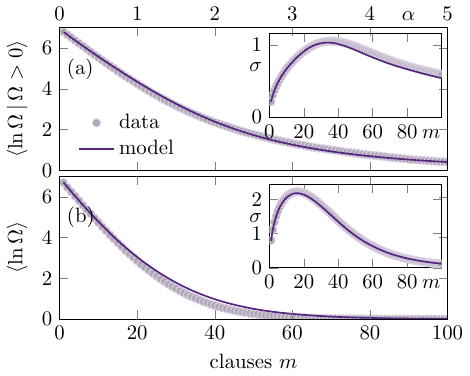}
    \caption{Mean $\langle\ \cdot\ \rangle$ and standard deviation $\sigma$ (inset) of the logarithmic number of satisfying assignments $\ln \Omega$ per row in \cref{fig:classical-approximation-illustration}. Data for random instances with $n=20$ sampled for $200$ instances and the predictions of the model in \cref{alg:stochastic} are shown. In (a) the rows are filtered to contain at least one satisfying assignment, while in (b) all rows are averaged.}
    \label{fig:shard-properties}
\end{figure}

For the Schmidt vectors to group satisfying assignments together, a variable assignment in one subsystem needs to have multiple possible satisfying completions in the other.
Given a row (or column) in the matrix in \cref{fig:classical-approximation-illustration}, the number of satisfying assignments $\Omega$ in it can be modeled as follows:
Each clause contains $k=3$ variables such that $i$ of them act on the row space with probability $p_i = \mathrm{Hypergeometric}(i; n_{\mathrm{row}}, n_{\mathrm{col}}, k)$ distributed hypergeometrically with $n_{\mathrm{row}}$ successes, $n_{\mathrm{col}}$ failures and $k$ tries. 
We assume even $n$ such that $n_{\mathrm{row}} = n_{\mathrm{col}} = n/2$.
For a given row with $i$ variables in the row space, the probability that this row already satisfies the drawn clause is $q_i = 2^{-i}$; if so, the number of satisfying assignments does not change.
If the row does not already satisfy the clause, on average, the fraction $f_i = 1 - 2^{i-k}$ of currently still satisfied assignments are ruled out by the clause.
Up to corrections to account for clauses with already drawn combinations of variables outlined in \cref{app:algo_details}, the following stochastic algorithm reproduces the distribution of $\Omega$: 
\begin{algorithm}[H]
    \caption{Stochastic model of \threesat{} groups}\label{alg:stochastic}
    \begin{algorithmic}[1]
        \State Initialize $\Omega = 2^{n/2}$
        \For{\textbf{each} clause $c$}
            \State Select $i$ according to probabilities $p_i$
            \State Do nothing with probability $q_i$
            \State Handle redundant cases, cf.~\cref{app:algo_details} 
            \State Update $\Omega \gets f_i \Omega$
        \EndFor
    \end{algorithmic}
\end{algorithm}

\Cref{fig:shard-properties} compares the model in \cref{alg:stochastic} to sampled data showing good agreement.
To simplify \cref{alg:stochastic}, we focus on a typical row and neglect the redundancy corrections. 
Then, the number of satisfying assignments $\Omega$ in such a row is obtained from
\begin{align*}
    \Omega = 2^{n/2} \prod_i f_i^{n_i} \implies \ln \Omega = \frac{\ln 2}{2} n + \sum_{i=0}^{k-1} q_i n_i \ln f_i,
\end{align*}
where $n_i$ is the number of times $i$ variables affected the row only, such that $\langle n_i \rangle = p_i m$; the probabilities $p_i \approx \binom{k}{i}/2^k$ are approximated binomially in the limit $n\rightarrow\infty$.
With this, the structural change close at the hardness transition can be both qualitatively and quantitatively corroborated.
Consider first the average number of satisfying assignments in a row $\langle \ln \Omega \rangle$ for the unconditioned case given by
\begin{align*}
    \langle \ln \Omega \rangle = \frac{\ln 2}{2} n + \alpha n \sum_{i=0}^{k-1} q_i p_i \ln f_i,
\end{align*}
which transitions from greater than zero to less than zero at
\begin{align*}
    \alpha^\star = -\ln 2 / 2\sum_{i=0}^{k-1} q_i p_i \ln f_i \approx 2.556.
\end{align*}
Hence, below $\alpha^\star$ each row and column on average contains at least one element, while above the threshold these occurrences get exceedingly sparse.
Then, above $\alpha^\star$ the state is formed by superpositions of the grouping of satisfying solutions illustrated in \cref{fig:entanglement-comparison-combinatorial-3sat}, while below $\alpha^\star$ the state is comprised of Schmidt vectors more akin to true quantum superpositions.

\begin{figure}
    \includegraphics{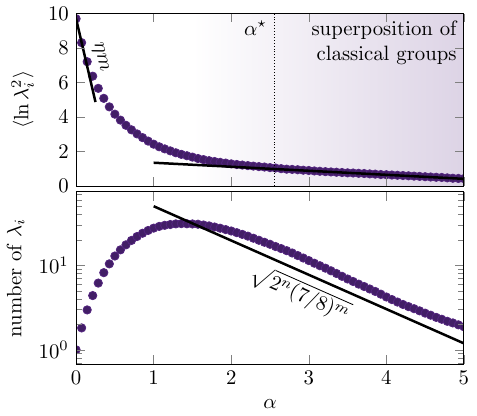}
    \caption{The expectation value of the logarithmic Schmidt values and the number of non-vanishing Schmidt values. In the beginning, the expectation values decline is dominated by the creation of superpositions across the cut, such that $\langle \ln \lambda_i^2 \rangle = \eta m$. At a later stage, the Schmidt vectors are superpositions of mostly frozen spins, as indicated by the slope. The number of Schmidt weights is roughly the estimate used for $\dim \mathcal L$ in the previous section, with the agreement getting better for larger systems (empirically, the constant is $2^{n+1}$ and not $2^n$). The data is collected for $1000$ instances with $n=14$.} \label{fig:schmidt-properties}
\end{figure}

As a diagnostic of this qualitative change, we consider the average logarithmic Schmidt value $\langle \ln \lambda_i \rangle$ in \cref{fig:schmidt-properties}, which is roughly related to the average number of satisfying assignments it represents.
While below the complexity transition ($\alpha \ll 1$) the Schmidt weights decay as qualitatively expected when increasingly more superpositions are formed, see \cref{app:initial_schmidt_values} for a derivation of the initial slope $\eta$, around the critical ratio $\alpha^\star$ the logarithmic Schmidt values decay with a much smaller slope.
In \cref{app:linear_growth_schmidt_values} we present an argument for why this linear slope is connected to classical superpositions.

Despite the grouping of logically correlated clauses becoming much less viable, the average number of Schmidt values used in \threesat{} states as plotted in \cref{fig:schmidt-properties} remains extensive in the remaining Hilbert size, cf.~the reservoir dimension in \cref{eq:reservoir_dimension}.
Thus, the states roughly live in an effective Hilbert space of dimension $\dim \mathcal L_m$, while not admitting to any further compression.
By this, the \gls{mps} is overwhelmed and approximability of the wavefunction becomes impossible.

\subsection{Implications for Imaginary Time Evolution}\label{sec:3}

In the previous section, we extensively discussed the breakdown mechanisms for the case in which the quantum state contains sufficient information to solve the $\sharp \P$ problem.
However, in systems deviating from the purely flat spectrum Hamiltonian, this ability is lost immediately.
Hence, in the quantum inspired algorithm involving imaginary time evolution, residues of the $\sharp \P$-hardness need to remain a challenge for the variational approach.

Returning to the imaginary time approach in \cref{eq:imagtimeevolution}, there each computational basis state was assigned an decay rate (energy) according to the number of clauses it violates.
As a simple approximation, consider the number of states with no violations $n_0$ and one violation $n_1$ in one row, cf.~\cref{fig:classical-approximation-illustration} and the previous discussion, which are expected to be 
\begin{align*}
    \langle n_0 \rangle &= \sqrt{2^n} \left(\frac{7}{8}\right)^m,&
    \langle n_1 \rangle &= \frac{m}{7} \langle n_0 \rangle,&
\end{align*}
for $n$ variables and $m = \alpha n$ clauses.
Thus, while $\langle n_0 \rangle$ is roughly constant close to the hardness transition $\alpha_\sharp$, the number of computational basis states with one violation scales as $\order{n}$.
Differentiating the desired states with no violations from those with few then involves either large imaginary time step to immediately suppress any solutions or large bond dimensions.
In the former case, the limit discussion in the previous section is obtained, in which \gls{mps} are restrained; the latter case gets computationally expensive due to the matrix operations involved~\cite{PAECKEL2019167998}.
While variable-to-clause ratios $\alpha$ are seemingly avoided in the \gls{itp}, they practically are not, as the time evolution~\cite{PAECKEL2019167998} needs to be implemented either locally~\cite{Suzuki1976,PhysRevLett.93.040502,PhysRevLett.93.076401} or applied in a sweep-like manner~\cite{PhysRevB.91.165112,PhysRevLett.107.070601,PhysRevB.94.165116}.

\section{Conclusion}\label{sec:conclusion}

In this paper, we studied the quantum resources required to solve satisfiability problems using imaginary time evolution.
While such a protocol requires only linear time in the size of the problem to solve it, we found that both entanglement and non-stabilizerness show characteristic bumps in imaginary time.
These bumps follow volume laws scaling linearly with problem size, such that the exponential hardness conjectured for satisfiability problems~\cite{EXPTIMECONJECTURE} translates to extensive quantum resource requirements in the protocol.
While this scaling of the incurred costs is independent of the problem configuration, i.e., for the satisfiability problem the ratio between variables and constraints, we found it to be the steepest not at the classically hardest point for the underlying satisfiability problem but instead when counting all the solutions is the hardest.

To substantiate the understanding of this resource bump, we considered a simplified protocol and developed a statistical model of the entanglement curves observed when additional constraints are enforced on a state.
This construction implies for \gls{mps} to be an efficient compression of exhaustive solution lists for verifying the solution of intermediate size problem instances.
Furthermore, this model corroborates previous findings regarding the position of the hardness transition for \sharpthreesat{}; tracking the critical ratio at which the entanglement bump occurs, we found it to approach $\alpha_\sharp\approx 2.595$ as was observed with purely classical solvers~\cite{BAILEY20071627}.
Close to the critical ratio, we investigated the structure of the basis states forming the \gls{mps} representation, finding a transition between a formation of genuine quantum transitions below the critical ratio to more classical exhaustive lists afterwards; right at the crossover finding an appropriate representation becomes the hardest.
Thus, the quantum perspective both is consistent with the heuristic determination of the hardness transition and provides a self contained picture of its origin.

\section*{Data Availability}

The data that support the findings of this article are openly available on Zenodo~\cite{pokart_2026_18743084}.

\begin{acknowledgements}
{T.P.} and {J.C.B.} acknowledge financial support from the Deutsche Forschungsgemeinschaft (DFG, German Research Foundation) through the Collaborative Research Centre SFB 1143, the Cluster of Excellence ct.qmat, and the DFG Project No.~419241108. 
F.P. acknowledges support from the Munich Quantum Valley, which is supported by the Bavarian state government with funds from the Hightech Agenda Bayern Plus, and the Munich Center for Quantum Science and Technology (MC-QST), supported by the DFG) under Germany’s Excellence Strategy Grant No.~EXC–2111–390814868.
The authors gratefully acknowledge the computing time made available to them on the high-performance computer at the NHR Center of TU Dresden. This center is jointly supported by the Federal Ministry of Education and Research and the state governments participating in the NHR.
\end{acknowledgements}

\appendix

\section{Temporal Location of the Entanglement Bump}\label{app:entanglement_scaling}

\begin{figure}
    \includegraphics{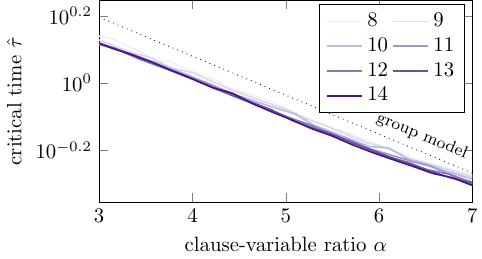}
    \caption{The imaginary time $\hat{\tau}$ at which the entanglement bump occurs and when averaged over $1000$ instances is universal and independent of the system size. }
    \label{fig:temporal-location-entanglement-bump}
\end{figure}

In \cref{fig:properties-entanglement-bump} the height $S^\star$ of the entanglement bump limiting the imaginary time evolution were shown.
Similarly, in \cref{fig:temporal-location-entanglement-bump} we show the expected time $\tau^\star$ at which the entanglement bump occurs, finding it to be independent of system size, again reflecting the expectation that the comparative hardness is dictated by $\alpha$.
The exponential scaling with $\alpha$, i.e., $\ln \tau^\star \propto \alpha$, is indicative of an intrinsic grouping within the state; that is: a state whose Schmidt vectors are built by grouping computational basis states with similar number of violations together also shows this behavior.
This will be shown in the following.

The scaling behavior of $\ln \tau \propto \alpha$ in \cref{fig:temporal-location-entanglement-bump} seems paradoxical at first, as the weights in the computational basis scale as $\exp(-v \tau)$ with $v$ the number of clauses a configuration violates.
Then, $v \propto \alpha$ would suggest that $\tau \propto 1/\alpha$.
This behavior can however be explained by considering a more sophisticated model, in which the Schmidt weights are also influenced by the number of states $\abs{X_v}$ which violate a given number of clauses $v \in \set{0, \ldots, m}$.
Then, the unnormalized Schmidt values if all states group into the same Schmidt vector is
\begin{align*}
    \tilde \lambda_v = \abs{X_v} \exp(-v \tau)
\end{align*}
with
\begin{align*}
    \abs{X_v} & = 2^n P(v), & P(v) & = \binom{m}{v} \left(\frac{1}{8}\right)^v \left(\frac{7}{8}\right)^{m-v} .
\end{align*}
First, we compute the normalization as
\begin{align*}
    \sum_{v=0}^m \tilde \lambda_v = 2^n \sum_{v=0}^m P(v) e^{-v\tau} = E[e^{-\tau X}] = 2^n (q - p e^{-\tau})^m.
\end{align*}
Then, $\lambda_v = \tilde \lambda_v / \sum_v \tilde \lambda_v$ are the normalized Schmidt values which yield the entanglement entropy $S = -\sum_v \lambda_v \ln \lambda_v$ in this simplified approach.
Then,
\begin{align*}
    S & = m \ln (q + p e^{-\tau})                                                                         \\
      & \qquad- \frac{1}{(q+pe^{-\tau})^m} \underbrace{\sum_v P(v) e^{-v\tau} \ln (P(v) e^{-v\tau})}_{Y}.
\end{align*}
This function has its maximum when $\partial_\tau S = 0$, which yields the condition
\begin{align*}
    0 & = \frac{m}{q + pe^{-\tau}} (-pe^{-\tau}) + \frac{m}{(q+pe^{-\tau})^{m+1}} (-pe^{-\tau}) Y \\
      & \qquad- \frac{Y'}{(q+pe^{-\tau})^m} \iff                                                  \\
    0 & = (q+pe^{-\tau})^m + Y + \frac{Y'}{m} (1 + qe^{\tau}/p).
\end{align*}
Lastly, we find the derivative $Y'$ to be
\begin{align*}
    Y' & = \underbrace{\sum_v (-v) P(v) e^{-v\tau} \ln (P(v) e^{-v\tau})}_{\approx -pmY }        \\
       & \qquad  + \underbrace{\sum_v (-v)e^{-v\tau}}_{\xrightarrow{m\rightarrow\infty} C(\tau)}.
\end{align*}
Now,
\begin{align*}
    0 & = m q^m + qmY (1 - e^\tau) + \underbrace{\frac{e^\tau (1 + qe^\tau/p)}{(1-e^\tau)^2}}_{\approx\ \mathrm{const.}}.
\end{align*}
We approximate $Y$ as
\begin{align*}
    Y & = \sum P(v) e^{-v\tau} \ln(P(v) e^{-v\tau})                     \\
      & = \sum e^{-v\tau} P(v) \ln P(v) - \tau \sum v P(v) \ln P(v)    \\
      & \approx e^{-pm\tau} H(m) - \tau p m H(m) \approx -\tau p m H(m),
\end{align*}
such that
\begin{align*}
     & q^m \propto \tau m^2 (1 - e^\tau) \approx \tau^2 m^2         \\
     & \qquad\implies \alpha \ln q = 2 \ln \tau + \mathrm{const.}
\end{align*}
Thus, a state in which computational basis states with similar violations group together already reproduces this behavior and we expect a slope of $\ln \tau = (\ln q / 2) \alpha$.

\section{Boolean Algebra Compression Decision Trees}\label{app:boolean_compression}

\begin{figure}
    \centering 
    \includegraphics{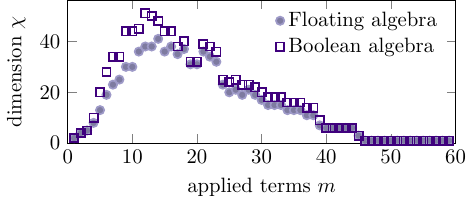}
    \caption{The dimension of the necessary Hilbert space of a bipartition for a \threesat{} instance with $n=12$ variables at different number of enforced clauses $m$ when using the Floating point algebra (the standard singular value decomposition used in \gls{mps}) compared to one basis using the Boolean algebra in \cref{eq:boolean_algebra}. Their qualitative similarity and quantitative closeness suggests that while not exact, the \gls{mps} vectors contain similar information as exact decision trees.}
    \label{fig:boolean_algebra}
\end{figure}

While the notion of associating the basis contained in the singular value decomposition involved in \gls{mps} with decision trees is not exact as non-Boolean floating point numbers are involved, it nevertheless serves as a good cartoon of the complex vectors.
Here, we investigate the discrepancy between this relaxed definition of decision trees and Boolean ones by adjusting the algebra used to build the basis, as building an independent set of sub-decision trees fundamentally only requires Gram Schmidt decomposition.
Hence, we may substitute the real algebra for a Boolean one, that is for two bitstrings $a$ and $b$ replace
\begin{align}
    (a \cdot b ) & \mapsto (a \Rightarrow b), &
    (a + b)      & \mapsto (a \lor b).
    \label{eq:boolean_algebra}
\end{align}
Here, the implication is to be understood element wise, that is for $a = a_1\ldots a_n$ and $b=b_1\ldots b_n$ we have that $a \Rightarrow b$ iff $a_i \Rightarrow b_i\ \forall i$.
Note that while the Gram Schmidt decomposition yields a full basis, it is not unique (neither the elements themselves nor their amount).
In \cref{fig:boolean_algebra}, we compare the dimension of a Hilbert space formed when bipartitioning a typical \threesat{} instance, finding the similar characteristics when the floating point algebra and Boolean algebra are used.
Thus, the \gls{mps} representation indeed contains (approximate) sub-decision trees.

\section{Calculations for the Independent Diagonal Model}\label{app:diagonal_model}

We compute the entanglement entropy for a bipartition of the system ($\ket{x}$) into left ($\ket{l_x}$) and right ($\ket{r_x}$) subsystems.
The reduced density matrix in the left system is then
\begin{align*}
    \rho = \trace_R \ketbra{\phi} = \sum_{x,y} \phi_x \phi_y (\braket{r_x}{r_y}) \ket{l_x}\bra{l_y}.
\end{align*}
The diagonal components of $\rho_{l_x l_x}$ are distributed as
\begin{align*}
    \rho_{l_x l_x} = \sum_{r} \phi_{l_x r}^2 \simeq \frac{F}{\dim \mathcal H} \times \dim \mathcal R,
\end{align*}
since for each of the $\dim \mathcal R$ candidates $x$ compatible with the left bitstring $l_x$, the probability to be unity is $F / \dim \mathcal H$.
For the off-diagonal components, we have
\begin{align*}
    \rho_{l_x l_y} = \sum_{r} \phi_{l_x r}\phi_{l_y r} \simeq \left(\frac{F}{\dim \mathcal H}\right)^2 \times \dim \mathcal R,
\end{align*}
which suppressed compared to the diagonal elements by a factor of
\begin{align*}
    F / \dim \mathcal H = 2^{(f-1)n} \xrightarrow[f\neq 1]{n\rightarrow \infty} 0
\end{align*}
and hence vanishes in the thermodynamic limit for any non-trivial point with filling $f \in (0, 1)$.
Therefore, we compute the entanglement entropy by considering only the diagonal entries~\cite{PhysRevLett.124.200602}, which are then modeled as Bernoulli random variables with probability $p = 2^{(f-1)n}$ drawn $N = \dim \mathcal R$ times.
The mean $\mu = Np$ and variance $\sigma^2 = N p (1-p)$ of this process are approximately equal, thus a Poisson distribution with parameter $\lambda = 2^{(f-1)n} \dim \mathcal R$ describes the distribution of diagonal 
entries in $\rho$.

From this, we obtain the expected entanglement entropy $\langle S \rangle = -\langle \sum_i \rho_{ii} \ln \rho_{ii} \rangle$  as
\begin{align*}
    \langle S \rangle = \dim \mathcal L \times\langle -\rho_{ii} \ln \rho_{ii} \rangle \equiv \langle s \rangle \dim \mathcal L .
\end{align*}
The average contribution per diagonal entry $\langle s \rangle$ can be obtained from the Poisson model as
\begin{align*}
    \langle s \rangle & = -\sum_{k=1}^\infty P(k) \frac{k}{F} \ln \frac{k}{F}                                         \\
                      & = -\frac{1}{F} \sum_{k=1}^\infty P(k) \left[k \ln k - k \ln F\right]                         \\
                      & = \frac{\lambda}{F} \left(\ln F + E_{X \sim \mathrm{Poisson}(\lambda)}[-\ln (X + 1)]\right).
\end{align*}
As $-\ln$ is a concave function, we apply Jensens inequality to it, namely that for $f$ convex it holds that $E[f(X)] \geq f(E[X])$.
The average entropy $\langle S \rangle$ is thus bounded from below by
\begin{align*}
    \langle S \rangle \geq \ln F - \ln (1 + \lambda) = fn - \ln (1 + 2^{(1/2 - f)n}).
\end{align*}
An upper bound may be established from Holevos theorem
\begin{align*}
    S\Big(\sum_{i}\lambda_i \rho_i\Big) \leq -\sum_i \lambda_i \ln \lambda_i + \sum_i \lambda_i S(\rho_i).
\end{align*}
Then,
\begin{align*}
    \langle S \rangle \leq \begin{cases}
                               fn,  & \text{if } f \leq n/2, \\
                               n/2, & \text{if } f > n/2.
                           \end{cases}
\end{align*}
This is also the thermodynamic limit $\lim_{n\rightarrow\infty}$ of the lower bound, such that for an infinite dimensional system the entanglement entropy grows linearly up until filling $f=n/2$ and then stays constant for $f < 1$.
At full-filling, it again vanishes.

\section{Deriving the Cancelation Correction Factor}\label{app:correlation_correction}
As illustrated in \cref{fig:graph}, some terms cancel each other, which requires some additional care.
The current number of triangles in the graph is
\begin{align*}
    t_{\mathrm{active}} = \frac{n_{\mathrm{active}}}{3}.
\end{align*}
In total, there can be at most
\begin{align*}
    t_{\mathrm{tot}} = \frac{8 n (n - 1) (n-2)}{3!}
\end{align*}
triangles, which is exactly the number of possible unique clauses with $n$ variables.
In order for a given triangle to generate additional entanglement, it should not match in the last variable with a previously drawn one, independent of the variables' polarity.
Hence, we may model this as follows: there are $N = t_{\mathrm{tot}}  / 2$ total possibilities to draw from in order to generate a collision and therefore in total $N^m$ possible combinations.
The number of drawings that do not generate a collision is
\begin{align*}
    \frac{N!}{(N - t_{\mathrm{active}})!},
\end{align*}
such that the triangle factor may be computed as
\begin{align*}
    f & = \frac{N!}{(N-A)! N^m}                                            \\
      & = \exp\left(\ln\Gamma(N+1) - \ln\Gamma(N-A+1) - m \ln N\right).
\end{align*}
Then, we adjust the entangling correlations as $C \rightarrow fC$.
In \cref{fig:entanglement-comparison-combinatorial-3sat} we show the entanglement properties obtained from this model compared to real \threesat{} instances.

\section{Details on Algorithm 1}\label{app:algo_details}

\begin{figure}
    \includegraphics{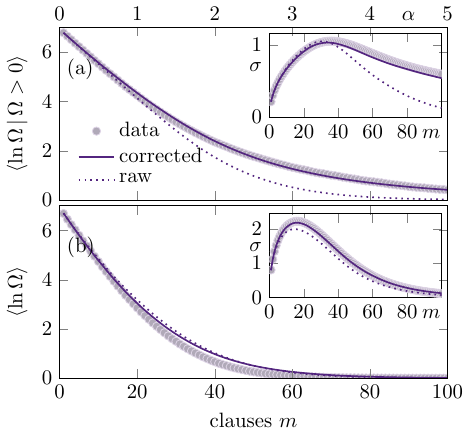}
    \caption{Same as \cref{fig:shard-properties} but also comparing the model in \cref{alg:stochastic} with correction (solid) and without corrections (dotted) in \cref{app:algo_details}. While the average data $\langle \ln \Omega \rangle$ is largely unchanged, the conditioned data $\langle \ln \Omega | \Omega > 0\rangle$ is improved.}\label{fig:shard_properties_detailed}
\end{figure}

Here, we provide details on how to correct for already drawn variable combinations in \cref{alg:stochastic}.
The necessity of this correction can be illustrated with the following example: suppose the \threesat{} instance contains the clauses $x_1 \lor x_2 \lor x_4$ and $x_1 \lor x_3 \lor x_4$.
Furthermore, let $x_1,x_2$ and $x_3$ constitute the row space and $x_4$ be a variable in the column space of the arrangement, cf.~the setup in \cref{fig:classical-approximation-illustration}.
Now, while the different configurations in the row space, i.e., $x_1\lor x_2$ for the first and $x_1 \lor x_3$ for the second clause, are already accounted for by skipping the clause for each sampled row with probability $q_i$ only dependent on the number of variables in the row space, the probability of the same variable, i.e., $x_4$ in our example, occurring multiple times in the column space would be neglected.
However, this repeated occurring reduces the effective fraction $f_i$ of states in a row being ruled unsatisfied.
Accurately accounting for this repetition is prone to overfitting, thus we resort to a very simple correction in the following, circumventing overfitting while nevertheless improving the accuracy of the model.
First, notice that in total, there are $N=2n(2n-2)(2n-4)$ possible clauses.
We now adjust the number of variables in the row space after already having seen $c$ clauses in the following way: 
\begin{enumerate}
    \item If $i=2$ there is one variable in the column space. Of the $N$ clauses, there are $n_1=3n_{\mathrm{row}}(n_{\mathrm{row}}-1)$ which contain the column variable at any of the three positions as well two suitable variables in the row space. Thus, with probability $p_1=(1-n_1/N)^{c-1}$ the current clause has no effect on the row and with probability $p_1$ the row vanishes (the clause containing the negated statement has been drawn).
    \item If $i=1$ there are two variables in the column space. There are $n_2=6 n_{\mathrm{row}}$ clauses which have the same selection of column variables and $n_3=3 n_{\mathrm{row}} 2(n_{\mathrm{col}}-1)$ clauses in which only one of the selected column variables occurs with the correct polarity. Thus, with $p_2=(1-n_2/N)^{c-1}$ the current clause has no effect, while we roll the probability $p_3=(1-n_3/N)^{c-1}$ twice and when successful decrease the number of column variables by $1$, i.e., $i \gets i+1$.
    \item If $i=0$ all three variables are in the column space. There are $n_4=6$ such clauses, thus with $p_4=(1-n_4/N)^{c-1}$ the clause has no effect; there are $n_5=6 \times 2(n_{\mathrm{col}}-1)$ clauses with two of the three variables occurring and thus rolling with $p_5=(1-n_5/N)^{c-1}$ thrice we decrease the number of column variables $i \gets i+1$; and the $n_6=3 \times 2(n_{\mathrm{col}}-1) \times 2(n_{\mathrm{col}}-2)$ cases in which one column variable already occurred are handled by rolling $p_6=(1-n_6/N)^{c-1}$ thrice and on success decreasing the number of column variables $i \gets i+2$.
\end{enumerate}
In \cref{fig:shard_properties_detailed} we amend \cref{fig:shard-properties} by the data obtained without the corrections above.
While the data not conditioned on non-vanished rows, i.e., $\langle \ln \Omega \rangle$, is largely unaffected, the corrections become very important for the conditioned expectation value $\langle \ln \Omega | \Omega > 0 \rangle$ where both mean and standard deviation improve drastically.

\section{Initial Schmidt Value Behavior}\label{app:initial_schmidt_values}

Initially, the Schmidt weight behavior is dominated by the build up of superpositions; thus, to determine this initial behavior, we consider a minimal example in which from the initial state in a three qubit system one computational basis state ($\ket{000}$ in the example) is removed.
Then, we make the following ansatz for the singular value decomposition
\begin{align*}
     & \ket{001} + \ket{010} + \ket{011} + \ket{100} + \ket{101} + \ket{110} + \ket{111}                 \\
     & = A (\cos \theta \ket{0} + \sin\theta \ket{1}) (\cos \alpha \ket{x} + \sin \alpha \ket{00})       \\
     & \quad + B (\sin \theta \ket{0} - \cos\theta \ket{1}) (\sin \alpha \ket{x} - \cos \alpha \ket{00})
\end{align*}
having defined $\ket{x} \equiv (\ket{01} + \ket{10} + \ket{11}) / \sqrt{3}$.
By comparing coefficients, we arrive at the system of equations:
\begin{align*}
    A \cos \theta \cos \alpha + B \sin \theta \sin \alpha & = \sqrt{3} \\
    A \sin \theta \cos \alpha - B \cos \theta \sin \alpha & = \sqrt{3} \\
    A \cos \theta \sin \alpha - B \sin \theta \cos \alpha & = 0        \\
    A \sin \theta \sin \alpha + B \cos \theta \cos \alpha & = 1.
\end{align*}
This may be solved to obtain the initial Schmidt weights
\begin{align*}
    A & = \sqrt{\frac{7 - \sqrt{37}}{2}} \approx 0.6772,                          \\
    B & = \frac{\sqrt{294} + \sqrt{222}}{12} \sqrt{7 - \sqrt{37}} \approx 2.5576.
\end{align*}
For the first clause, it either (i) acts exclusively in the left or right subsystem or (ii) concerns variables in both subsystems.
Only in case (ii) Schmidt values are generated, while case (i) leaves the imaginary cut untouched.
Let $l$ ($r$) be the number of variables in the left (right) subsystem, then the probability to act exclusively in the left or right subsystems is
\begin{align*}
    p = \frac{k (k - 1) (k - 2) + r (r-1) (r-2)}{n(n-1)(n-2)}.
\end{align*}
Now, with probability $p$ no Schmidt pair is generated, but the norm of the vector is simply reduced by $7/8$, and with probability $1-p$ a Schmidt pair is generated having the weights as above.
The logarithmic Schmidt value thus behaves like
\begin{align*}
    \ln \lambda_i^2 & \rightarrow \frac{p \ln (\frac{7}{8} \lambda_i^2) + (1-p) [\ln (\frac{A^2}{8}\lambda_i^2) + \ln (\frac{B^2}{8} \lambda_i^2)]}{p + 2 (1-p)} \\
                     & = \ln \lambda_i^2 + \eta,
\end{align*}
with 
\begin{align*}
    \eta \equiv \frac{p \ln (7/8) + (1-p) [\ln (A^2/8) + \ln (B^2/8)]}{p + 2(1-p)},
\end{align*}
which yields the initial slope of the expecation value as $\langle \ln \lambda_i^2 \rangle \propto \eta m$, cf.\ \cref{fig:schmidt-properties} for an empirical validation.

\section{On the Linear Growth of Logarithmic Schmidt Weights}\label{app:linear_growth_schmidt_values}

In \cref{fig:schmidt-properties} we showed numerical evidence that the logarithmic Schmidt weights $\ln \lambda$ decay approximately linearly after the complexity transition at $\alpha^\star$.
This may be explained by the following model: when enforcing an additional clause, the classical solution subspace described by the Schmidt vectors may (i) completely vanish, thus have no influence on the new mean, or (ii) remain at least partially intact and with some probability $p$ remain unaffected or with probability $1-p$ lose some fraction $q$ of its solution space.
Then, the remaining  logarithmic Schmidt weights transform as 
\begin{align*}
    &\ln \lambda \rightarrow p \ln \lambda + (1-p) \ln \sqrt{q} \lambda \\
    &\qquad\implies \ln \lambda \rightarrow \ln \lambda + (1-p) \ln \sqrt{q}.
\end{align*}
Hence we expect a constant slope of $(1-p) \ln \sqrt{q}<0$ once the Schmidt weights describe somewhat well defined classical configurations of solutions.


\phantomsection
\addcontentsline{toc}{chapter}{Bibliography}
\bibliography{literature.bib}

@article{PhysRevX.14.041029,
  title     = {Opening the Black Box inside Grover's Algorithm},
  author    = {Stoudenmire, E. M. and Waintal, Xavier},
  journal   = {Phys. Rev. X},
  volume    = {14},
  issue     = {4},
  pages     = {041029},
  numpages  = {22},
  year      = {2024},
  month     = {Nov},
  publisher = {American Physical Society},
  doi       = {10.1103/PhysRevX.14.041029},
  url       = {https://link.aps.org/doi/10.1103/PhysRevX.14.041029}
}

@article{GAREY1976237,
  title   = {Some simplified NP-complete graph problems},
  journal = {Theoretical Computer Science},
  volume  = {1},
  number  = {3},
  pages   = {237-267},
  year    = {1976},
  issn    = {0304-3975},
  doi     = {https://doi.org/10.1016/0304-3975(76)90059-1},
  url     = {https://www.sciencedirect.com/science/article/pii/0304397576900591},
  author  = {M.R. Garey and D.S. Johnson and L. Stockmeyer}
}

@article{BAILEY20071627,
  title    = {Phase transitions of PP-complete satisfiability problems},
  journal  = {Discrete Applied Mathematics},
  volume   = {155},
  number   = {12},
  pages    = {1627-1639},
  year     = {2007},
  note     = {SAT 2001, the Fourth International Symposium on the Theory and Applications of Satisfiability Testing},
  issn     = {0166-218X},
  doi      = {https://doi.org/10.1016/j.dam.2006.09.014},
  url      = {https://www.sciencedirect.com/science/article/pii/S0166218X06004665},
  author   = {Delbert D. Bailey and Víctor Dalmau and Phokion G. Kolaitis}
}

@inproceedings{pmlr-v97-wang19e,
  title     = {{SATN}et: Bridging deep learning and logical reasoning using a differentiable satisfiability solver},
  author    = {Wang, Po-Wei and Donti, Priya and Wilder, Bryan and Kolter, Zico},
  booktitle = {Proceedings of the 36th International Conference on Machine Learning},
  pages     = {6545--6554},
  year      = {2019},
  editor    = {Chaudhuri, Kamalika and Salakhutdinov, Ruslan},
  volume    = {97},
  series    = {Proceedings of Machine Learning Research},
  month     = {09--15 Jun},
  publisher = {PMLR},
  url       = {https://proceedings.mlr.press/v97/wang19e.html}
}

@inproceedings{10.1007/978-3-319-40970-2_15,
  title     = {Solving and Verifying the Boolean Pythagorean Triples Problem via Cube-and-Conquer},
  booktitle = {Theory and Applications of Satisfiability Testing -- {{SAT}} 2016},
  author    = {Heule, Marijn J. H. and Kullmann, Oliver and Marek, Victor W.},
  editor    = {Creignou, Nadia and Le Berre, Daniel},
  year      = {2016},
  pages     = {228--245},
  publisher = {Springer International Publishing},
  address   = {Cham},
  isbn      = {978-3-319-40970-2},
  doi       = {10.1007/978-3-319-40970-2_15}
}

@inproceedings{10.1007/978-3-030-42266-0_6,
  title     = {Comparing Integer Linear Programming to {{SAT-solving}} for Hard Problems in Computational and Systems Biology},
  booktitle = {Algorithms for Computational Biology},
  author    = {Brown, Hannah and Zuo, Lei and Gusfield, Dan},
  editor    = {Martín-Vide, Carlos and Vega-Rodríguez, Miguel A. and Wheeler, Travis},
  year      = {2020},
  pages     = {63--76},
  publisher = {Springer International Publishing},
  location  = {Cham},
  isbn      = {978-3-030-42266-0},
  doi       = {10.1007/978-3-030-42266-0_6}
}

@inproceedings{SATUNSATTRANSITION2,
  author    = {Mitchell, David and Selman, Bart and Levesque, Hector},
  title     = {Hard and easy distributions of SAT problems},
  year      = {1992},
  isbn      = {0262510634},
  publisher = {AAAI Press},
  booktitle = {Proceedings of the Tenth National Conference on Artificial Intelligence},
  pages     = {459–465},
  numpages  = {7},
  location  = {San Jose, California},
  series    = {AAAI'92},
  doi       = {10.5555/1867135.1867206}
}

@inproceedings{SATUNSATTRANSITION1,
  author    = {Cheeseman, Peter and Kanefsky, Bob and Taylor, William M.},
  title     = {Where the really hard problems are},
  year      = {1991},
  isbn      = {1558601600},
  publisher = {Morgan Kaufmann Publishers Inc.},
  address   = {San Francisco, CA, USA},
  booktitle = {Proceedings of the 12th International Joint Conference on Artificial Intelligence - Volume 1},
  pages     = {331–337},
  numpages  = {7},
  location  = {Sydney, New South Wales, Australia},
  series    = {IJCAI'91},
  doi       = {10.5555/1631171.1631221}
}

@article{REPLICAMETHOD1,
  author  = {Monasson, R{\'e}mi
             and Zecchina, Riccardo
             and Kirkpatrick, Scott
             and Selman, Bart
             and Troyansky, Lidror},
  title   = {Determining computational complexity from characteristic `phase transitions'},
  journal = {Nature},
  year    = {1999},
  month   = {Jul},
  day     = {01},
  volume  = {400},
  number  = {6740},
  pages   = {133-137},
  issn    = {1476-4687},
  doi     = {10.1038/22055},
  url     = {https://doi.org/10.1038/22055}
}

@article{PhysRevLett.124.200602,
  title     = {Multifractality Meets Entanglement: Relation for Nonergodic Extended States},
  author    = {De Tomasi, Giuseppe and Khaymovich, Ivan M.},
  journal   = {Phys. Rev. Lett.},
  volume    = {124},
  issue     = {20},
  pages     = {200602},
  numpages  = {7},
  year      = {2020},
  month     = {May},
  publisher = {American Physical Society},
  doi       = {10.1103/PhysRevLett.124.200602},
  url       = {https://link.aps.org/doi/10.1103/PhysRevLett.124.200602}
}

@article{PhysRevA.74.060304,
  title     = {Adiabatic quantum algorithms as quantum phase transitions: First versus second order},
  author    = {Sch\"utzhold, Ralf and Schaller, Gernot},
  journal   = {Phys. Rev. A},
  volume    = {74},
  issue     = {6},
  pages     = {060304},
  numpages  = {4},
  year      = {2006},
  month     = {Dec},
  publisher = {American Physical Society},
  doi       = {10.1103/PhysRevA.74.060304},
  url       = {https://link.aps.org/doi/10.1103/PhysRevA.74.060304}
}

@inproceedings{GROVERSEARCH,
  author    = {Grover, Lov K.},
  title     = {A fast quantum mechanical algorithm for database search},
  year      = {1996},
  isbn      = {0897917855},
  publisher = {Association for Computing Machinery},
  address   = {New York, NY, USA},
  url       = {https://doi.org/10.1145/237814.237866},
  doi       = {10.1145/237814.237866},
  booktitle = {Proceedings of the Twenty-Eighth Annual ACM Symposium on Theory of Computing},
  pages     = {212–219},
  numpages  = {8},
  location  = {Philadelphia, Pennsylvania, USA},
  series    = {STOC '96}
}

@misc{ADIABATICQUANTUMCOMPUTING1,
  title         = {Quantum Computation by Adiabatic Evolution},
  author        = {Edward Farhi and Jeffrey Goldstone and Sam Gutmann and Michael Sipser},
  year          = {2000},
  eprint        = {quant-ph/0001106},
  archiveprefix = {arXiv},
  primaryclass  = {quant-ph},
  url           = {https://arxiv.org/abs/quant-ph/0001106}
}

@article{ADIABATICQUANTUMCOMPUTING2,
  author  = {Edward Farhi  and Jeffrey Goldstone  and Sam Gutmann  and Joshua Lapan  and Andrew Lundgren  and Daniel Preda },
  title   = {A Quantum Adiabatic Evolution Algorithm Applied to Random Instances of an NP-Complete Problem},
  journal = {Science},
  volume  = {292},
  number  = {5516},
  pages   = {472-475},
  year    = {2001},
  doi     = {10.1126/science.1057726}
}

@article{PhysRevA.83.012309,
  title     = {Classical and quantum annealing in the median of three-satisfiability},
  author    = {Neuhaus, T. and Peschina, M. and Michielsen, K. and De Raedt, H.},
  journal   = {Phys. Rev. A},
  volume    = {83},
  issue     = {1},
  pages     = {012309},
  numpages  = {5},
  year      = {2011},
  month     = {Jan},
  publisher = {American Physical Society},
  doi       = {10.1103/PhysRevA.83.012309},
  url       = {https://link.aps.org/doi/10.1103/PhysRevA.83.012309}
}

@article{ADIABATICQUANTUMCOMPUTING3,
  author     = {Farhi, Edward and Goldston, Jeffrey and Gosset, David and Gutmann, Sam and Meyer, Harvey B. and Shor, Peter},
  title      = {Quantum adiabatic algorithms, small gaps, and different paths},
  year       = {2011},
  issue_date = {March 2011},
  publisher  = {Rinton Press, Incorporated},
  address    = {Paramus, NJ},
  volume     = {11},
  number     = {3},
  issn       = {1533-7146},
  journal    = {Quantum Info. Comput.},
  month      = mar,
  pages      = {181–214},
  numpages   = {34},
  doi        = {10.5555/2011395.2011396}
}

@article{PhysRevA.61.052311,
  title     = {Quantum search heuristics},
  author    = {Hogg, Tad},
  journal   = {Phys. Rev. A},
  volume    = {61},
  issue     = {5},
  pages     = {052311},
  numpages  = {7},
  year      = {2000},
  month     = {Apr},
  publisher = {American Physical Society},
  doi       = {10.1103/PhysRevA.61.052311},
  url       = {https://link.aps.org/doi/10.1103/PhysRevA.61.052311}
}

@article{PRXQuantum.5.030348,
  title     = {Solving Boolean Satisfiability Problems With The Quantum Approximate Optimization Algorithm},
  author    = {Boulebnane, Sami and Montanaro, Ashley},
  journal   = {PRX Quantum},
  volume    = {5},
  issue     = {3},
  pages     = {030348},
  numpages  = {32},
  year      = {2024},
  month     = {Sep},
  publisher = {American Physical Society},
  doi       = {10.1103/PRXQuantum.5.030348},
  url       = {https://link.aps.org/doi/10.1103/PRXQuantum.5.030348}
}

@article{PhysRevApplied.17.024052,
  title     = {Spintronics-compatible Approach to Solving Maximum-Satisfiability Problems with Probabilistic Computing, Invertible Logic, and Parallel Tempering},
  author    = {Grimaldi, Andrea and S\'anchez-Tejerina, Luis and Anjum Aadit, Navid and Chiappini, Stefano and Carpentieri, Mario and Camsari, Kerem and Finocchio, Giovanni},
  journal   = {Phys. Rev. Appl.},
  volume    = {17},
  issue     = {2},
  pages     = {024052},
  numpages  = {10},
  year      = {2022},
  month     = {Feb},
  publisher = {American Physical Society},
  doi       = {10.1103/PhysRevApplied.17.024052},
  url       = {https://link.aps.org/doi/10.1103/PhysRevApplied.17.024052}
}

@article{PhysRevApplied.23.034031,
  title     = {Parallel tempering--inspired distributed binary optimization with in-memory computing},
  author    = {Zhang, Xiangyi and Valiante, Elisabetta and Noori, Moslem and Yang, Chan-Woo and Rozada, Ignacio and B\"ohm, Fabian and Van Vaerenbergh, Thomas and Pedretti, Giacomo and Mohseni, Masoud and Beausoleil, Raymond},
  journal   = {Phys. Rev. Appl.},
  volume    = {23},
  issue     = {3},
  pages     = {034031},
  numpages  = {15},
  year      = {2025},
  month     = {Mar},
  publisher = {American Physical Society},
  doi       = {10.1103/PhysRevApplied.23.034031},
  url       = {https://link.aps.org/doi/10.1103/PhysRevApplied.23.034031}
}

@inproceedings{10.1007/978-3-030-14082-3_3,
  author    = {Gabor, Thomas
               and Zielinski, Sebastian
               and Feld, Sebastian
               and Roch, Christoph
               and Seidel, Christian
               and Neukart, Florian
               and Galter, Isabella
               and Mauerer, Wolfgang
               and Linnhoff-Popien, Claudia},
  editor    = {Feld, Sebastian
               and Linnhoff-Popien, Claudia},
  title     = {Assessing Solution Quality of 3SAT on a Quantum Annealing Platform},
  booktitle = {Quantum Technology and Optimization Problems},
  year      = {2019},
  publisher = {Springer International Publishing},
  address   = {Cham},
  pages     = {23--35},
  isbn      = {978-3-030-14082-3},
  doi       = {10.1007/978-3-030-14082-3_3}
}

@article{PARALLELTEMPERING2,
  doi       = {10.1209/0295-5075/19/6/002},
  url       = {https://dx.doi.org/10.1209/0295-5075/19/6/002},
  year      = {1992},
  month     = {jul},
  publisher = {},
  volume    = {19},
  number    = {6},
  pages     = {451},
  author    = {E. Marinari and G. Parisi},
  title     = {Simulated Tempering: A New Monte Carlo Scheme},
  journal   = {Europhysics Letters}
}

@article{PARALLELTEMPERING1,
  title     = {Replica Monte Carlo Simulation of Spin-Glasses},
  author    = {Swendsen, Robert H. and Wang, Jian-Sheng},
  journal   = {Phys. Rev. Lett.},
  volume    = {57},
  issue     = {21},
  pages     = {2607--2609},
  numpages  = {0},
  year      = {1986},
  month     = {Nov},
  publisher = {American Physical Society},
  doi       = {10.1103/PhysRevLett.57.2607},
  url       = {https://link.aps.org/doi/10.1103/PhysRevLett.57.2607}
}

@article{SIMULATEDANNEALING,
  author  = {S. Kirkpatrick  and C. D. Gelatt  and M. P. Vecchi },
  title   = {Optimization by Simulated Annealing},
  journal = {Science},
  volume  = {220},
  number  = {4598},
  pages   = {671-680},
  year    = {1983},
  doi     = {10.1126/science.220.4598.671},
  url     = {https://www.science.org/doi/abs/10.1126/science.220.4598.671}
}

@misc{huynh2023quantuminspiredmachinelearningsurvey,
  title         = {Quantum-Inspired Machine Learning: a Survey},
  author        = {Larry Huynh and Jin Hong and Ajmal Mian and Hajime Suzuki and Yanqiu Wu and Seyit Camtepe},
  year          = {2023},
  eprint        = {2308.11269},
  archiveprefix = {arXiv},
  primaryclass  = {cs.LG},
  url           = {https://arxiv.org/abs/2308.11269}
}

@article{Alvarez-Alvarado2021,
  author  = {Alvarez-Alvarado, Manuel S.
             and Alban-Chac{\'o}n, Francisco E.
             and Lamilla-Rubio, Erick A.
             and Rodr{\'i}guez-Gallegos, Carlos D.
             and Vel{\'a}squez, Washington},
  title   = {Three novel quantum-inspired swarm optimization algorithms using different bounded potential fields},
  journal = {Scientific Reports},
  year    = {2021},
  month   = {Jun},
  day     = {02},
  volume  = {11},
  number  = {1},
  pages   = {11655},
  issn    = {2045-2322},
  doi     = {10.1038/s41598-021-90847-7},
  url     = {https://doi.org/10.1038/s41598-021-90847-7}
}

@article{PRXQuantum.5.010308,
  title     = {Efficient Tensor Network Simulation of IBM's Eagle Kicked Ising Experiment},
  author    = {Tindall, Joseph and Fishman, Matthew and Stoudenmire, E. Miles and Sels, Dries},
  journal   = {PRX Quantum},
  volume    = {5},
  issue     = {1},
  pages     = {010308},
  numpages  = {16},
  year      = {2024},
  month     = {Jan},
  publisher = {American Physical Society},
  doi       = {10.1103/PRXQuantum.5.010308},
  url       = {https://link.aps.org/doi/10.1103/PRXQuantum.5.010308}
}

@misc{pednault2019leveragingsecondarystoragesimulate,
  title         = {Leveraging Secondary Storage to Simulate Deep 54-qubit Sycamore Circuits},
  author        = {Edwin Pednault and John A. Gunnels and Giacomo Nannicini and Lior Horesh and Robert Wisnieff},
  year          = {2019},
  eprint        = {1910.09534},
  archiveprefix = {arXiv},
  primaryclass  = {quant-ph},
  url           = {https://arxiv.org/abs/1910.09534}
}

@article{PhysRevX.10.041038,
  title     = {What Limits the Simulation of Quantum Computers?},
  author    = {Zhou, Yiqing and Stoudenmire, E. Miles and Waintal, Xavier},
  journal   = {Phys. Rev. X},
  volume    = {10},
  issue     = {4},
  pages     = {041038},
  numpages  = {15},
  year      = {2020},
  month     = {Nov},
  publisher = {American Physical Society},
  doi       = {10.1103/PhysRevX.10.041038},
  url       = {https://link.aps.org/doi/10.1103/PhysRevX.10.041038}
}

@misc{garciasaez2011exacttensornetwork3sat,
  title         = {An exact tensor network for the 3SAT problem},
  author        = {A. Garcia-Saez and J. I. Latorre},
  year          = {2011},
  eprint        = {1105.3201},
  archiveprefix = {arXiv},
  primaryclass  = {quant-ph},
  url           = {https://arxiv.org/abs/1105.3201}
}

@article{10.21468/SciPostPhys.7.5.060,
  title     = {{Fast counting with tensor networks}},
  author    = {Stefanos Kourtis and Claudio Chamon and Eduardo R. Mucciolo and Andrei E. Ruckenstein},
  journal   = {SciPost Phys.},
  volume    = {7},
  pages     = {060},
  year      = {2019},
  publisher = {SciPost},
  doi       = {10.21468/SciPostPhys.7.5.060},
  url       = {https://scipost.org/10.21468/SciPostPhys.7.5.060}
}

@inproceedings{10.1145/800157.805047,
  author    = {Cook, Stephen A.},
  title     = {The complexity of theorem-proving procedures},
  year      = {1971},
  isbn      = {9781450374644},
  publisher = {Association for Computing Machinery},
  address   = {New York, NY, USA},
  url       = {https://doi.org/10.1145/800157.805047},
  doi       = {10.1145/800157.805047},
  booktitle = {Proceedings of the Third Annual ACM Symposium on Theory of Computing},
  pages     = {151–158},
  numpages  = {8},
  location  = {Shaker Heights, Ohio, USA},
  series    = {STOC '71}
}

@inbook{Karp1972,
  author    = {Karp, Richard M.},
  editor    = {Miller, Raymond E.
               and Thatcher, James W.
               and Bohlinger, Jean D.},
  title     = {Reducibility among Combinatorial Problems},
  booktitle = {Complexity of Computer Computations: Proceedings of a symposium on the Complexity of Computer Computations, held March 20--22, 1972, at the IBM Thomas J. Watson Research Center, Yorktown Heights, New York, and sponsored by the Office of Naval Research, Mathematics Program, IBM World Trade Corporation, and the IBM Research Mathematical Sciences Department},
  year      = {1972},
  publisher = {Springer US},
  address   = {Boston, MA},
  pages     = {85--103},
  isbn      = {978-1-4684-2001-2},
  doi       = {10.1007/978-1-4684-2001-2_9},
  url       = {https://doi.org/10.1007/978-1-4684-2001-2_9}
}

@inproceedings{EXPTIMECONJECTURE,
  author    = {Impagliazzo, R. and Paturi, R.},
  booktitle = {Proceedings. Fourteenth Annual IEEE Conference on Computational Complexity (Formerly: Structure in Complexity Theory Conference) (Cat.No.99CB36317)},
  title     = {Complexity of k-SAT},
  year      = {1999},
  volume    = {},
  number    = {},
  pages     = {237-240},
  doi       = {10.1109/CCC.1999.766282}
}

@misc{watrous2008quantumcomputationalcomplexity,
  title         = {Quantum Computational Complexity},
  author        = {John Watrous},
  year          = {2008},
  eprint        = {0804.3401},
  archiveprefix = {arXiv},
  primaryclass  = {quant-ph},
  url           = {https://arxiv.org/abs/0804.3401}
}

@inproceedings{10.5555/3157382.3157634,
  author    = {Stoudenmire, E. M. and Schwab, David J.},
  title     = {Supervised learning with tensor networks},
  year      = {2016},
  isbn      = {9781510838819},
  publisher = {Curran Associates Inc.},
  address   = {Red Hook, NY, USA},
  booktitle = {Proceedings of the 30th International Conference on Neural Information Processing Systems},
  pages     = {4806–4814},
  numpages  = {9},
  location  = {Barcelona, Spain},
  series    = {NIPS'16},
  doi       = {10.5555/3157382.3157634}
}

@article{PhysRevResearch.6.043077,
  title     = {Temporal entanglement barriers in dual-unitary Clifford circuits with measurements},
  author    = {Yao, Jiangtian and Claeys, Pieter W.},
  journal   = {Phys. Rev. Res.},
  volume    = {6},
  issue     = {4},
  pages     = {043077},
  numpages  = {28},
  year      = {2024},
  month     = {Oct},
  publisher = {American Physical Society},
  doi       = {10.1103/PhysRevResearch.6.043077},
  url       = {https://link.aps.org/doi/10.1103/PhysRevResearch.6.043077}
}

@article{PhysRevLett.93.040502,
  title     = {Efficient Simulation of One-Dimensional Quantum Many-Body Systems},
  author    = {Vidal, Guifr\'e},
  journal   = {Phys. Rev. Lett.},
  volume    = {93},
  issue     = {4},
  pages     = {040502},
  numpages  = {4},
  year      = {2004},
  month     = {Jul},
  publisher = {American Physical Society},
  doi       = {10.1103/PhysRevLett.93.040502},
  url       = {https://link.aps.org/doi/10.1103/PhysRevLett.93.040502}
}

@article{Suzuki1976,
  author  = {Suzuki, Masuo},
  title   = {Generalized Trotter's formula and systematic approximants of exponential operators and inner derivations with applications to many-body problems},
  journal = {Communications in Mathematical Physics},
  year    = {1976},
  month   = {Jun},
  day     = {01},
  volume  = {51},
  number  = {2},
  pages   = {183-190},
  issn    = {1432-0916},
  doi     = {10.1007/BF01609348},
  url     = {https://doi.org/10.1007/BF01609348}
}

@article{PhysRevLett.93.076401,
  title     = {Real-Time Evolution Using the Density Matrix Renormalization Group},
  author    = {White, Steven R. and Feiguin, Adrian E.},
  journal   = {Phys. Rev. Lett.},
  volume    = {93},
  issue     = {7},
  pages     = {076401},
  numpages  = {4},
  year      = {2004},
  month     = {Aug},
  publisher = {American Physical Society},
  doi       = {10.1103/PhysRevLett.93.076401},
  url       = {https://link.aps.org/doi/10.1103/PhysRevLett.93.076401}
}

@article{PhysRevB.91.165112,
  title     = {Time-evolving a matrix product state with long-ranged interactions},
  author    = {Zaletel, Michael P. and Mong, Roger S. K. and Karrasch, Christoph and Moore, Joel E. and Pollmann, Frank},
  journal   = {Phys. Rev. B},
  volume    = {91},
  issue     = {16},
  pages     = {165112},
  numpages  = {8},
  year      = {2015},
  month     = {Apr},
  publisher = {American Physical Society},
  doi       = {10.1103/PhysRevB.91.165112},
  url       = {https://link.aps.org/doi/10.1103/PhysRevB.91.165112}
}

@article{PhysRevLett.107.070601,
  title     = {Time-Dependent Variational Principle for Quantum Lattices},
  author    = {Haegeman, Jutho and Cirac, J. Ignacio and Osborne, Tobias J. and Pi\ifmmode \check{z}\else \v{z}\fi{}orn, Iztok and Verschelde, Henri and Verstraete, Frank},
  journal   = {Phys. Rev. Lett.},
  volume    = {107},
  issue     = {7},
  pages     = {070601},
  numpages  = {5},
  year      = {2011},
  month     = {Aug},
  publisher = {American Physical Society},
  doi       = {10.1103/PhysRevLett.107.070601},
  url       = {https://link.aps.org/doi/10.1103/PhysRevLett.107.070601}
}

@article{PhysRevB.94.165116,
  title     = {Unifying time evolution and optimization with matrix product states},
  author    = {Haegeman, Jutho and Lubich, Christian and Oseledets, Ivan and Vandereycken, Bart and Verstraete, Frank},
  journal   = {Phys. Rev. B},
  volume    = {94},
  issue     = {16},
  pages     = {165116},
  numpages  = {10},
  year      = {2016},
  month     = {Oct},
  publisher = {American Physical Society},
  doi       = {10.1103/PhysRevB.94.165116},
  url       = {https://link.aps.org/doi/10.1103/PhysRevB.94.165116}
}

@article{PAECKEL2019167998,
  title    = {Time-evolution methods for matrix-product states},
  journal  = {Annals of Physics},
  volume   = {411},
  pages    = {167998},
  year     = {2019},
  issn     = {0003-4916},
  doi      = {https://doi.org/10.1016/j.aop.2019.167998},
  url      = {https://www.sciencedirect.com/science/article/pii/S0003491619302532},
  author   = {Sebastian Paeckel and Thomas Köhler and Andreas Swoboda and Salvatore R. Manmana and Ulrich Schollwöck and Claudius Hubig}
}

@inproceedings{Shor94,
  author    = {Shor, P.W.},
  booktitle = {Proceedings 35th Annual Symposium on Foundations of Computer Science},
  title     = {Algorithms for quantum computation: discrete logarithms and factoring},
  year      = {1994},
  volume    = {},
  number    = {},
  pages     = {124-134},
  doi       = {10.1109/SFCS.1994.365700}
}

@misc{hu2025asymptoticexceptionalsteadystates,
  title         = {Asymptotic Exceptional Steady States in Dissipative Dynamics},
  author        = {Yu-Min Hu and Jan Carl Budich},
  year          = {2025},
  eprint        = {2504.02937},
  archiveprefix = {arXiv},
  primaryclass  = {quant-ph},
  url           = {https://arxiv.org/abs/2504.02937}
}

@book{Aaronson_2013,
  place     = {Cambridge},
  title     = {Quantum Computing since Democritus},
  publisher = {Cambridge University Press},
  author    = {Aaronson, Scott},
  year      = {2013},
  doi       = {10.1017/CBO9780511979309}
}

@incollection{aaronson_6555741,
  author    = {Aaronson, Scott},
  isbn      = {9780262312677},
  title     = {Why Philosophers Should Care about Computational Complexity},
  booktitle = {Computability: Turing, Gödel, Church, and Beyond},
  publisher = {The MIT Press},
  year      = {2013},
  month     = {06},
  doi       = {10.7551/mitpress/8009.003.0011},
  url       = {https://doi.org/10.7551/mitpress/8009.003.0011},
  eprint    = {1108.1791}
}

@article{RevModPhys.90.015002,
  title     = {Adiabatic quantum computation},
  author    = {Albash, Tameem and Lidar, Daniel A.},
  journal   = {Rev. Mod. Phys.},
  volume    = {90},
  issue     = {1},
  pages     = {015002},
  numpages  = {64},
  year      = {2018},
  month     = {Jan},
  publisher = {American Physical Society},
  doi       = {10.1103/RevModPhys.90.015002},
  url       = {https://link.aps.org/doi/10.1103/RevModPhys.90.015002}
}

@article{RevModPhys.94.015004,
  title     = {Noisy intermediate-scale quantum algorithms},
  author    = {Bharti, Kishor and Cervera-Lierta, Alba and Kyaw, Thi Ha and Haug, Tobias and Alperin-Lea, Sumner and Anand, Abhinav and Degroote, Matthias and Heimonen, Hermanni and Kottmann, Jakob S. and Menke, Tim and Mok, Wai-Keong and Sim, Sukin and Kwek, Leong-Chuan and Aspuru-Guzik, Al\'an},
  journal   = {Rev. Mod. Phys.},
  volume    = {94},
  issue     = {1},
  pages     = {015004},
  numpages  = {69},
  year      = {2022},
  month     = {Feb},
  publisher = {American Physical Society},
  doi       = {10.1103/RevModPhys.94.015004},
  url       = {https://link.aps.org/doi/10.1103/RevModPhys.94.015004}
}

@article{TILLY20221,
  title    = {The Variational Quantum Eigensolver: A review of methods and best practices},
  journal  = {Physics Reports},
  volume   = {986},
  pages    = {1-128},
  year     = {2022},
  note     = {The Variational Quantum Eigensolver: a review of methods and best practices},
  issn     = {0370-1573},
  doi      = {https://doi.org/10.1016/j.physrep.2022.08.003},
  url      = {https://www.sciencedirect.com/science/article/pii/S0370157322003118},
  author   = {Jules Tilly and Hongxiang Chen and Shuxiang Cao and Dario Picozzi and Kanav Setia and Ying Li and Edward Grant and Leonard Wossnig and Ivan Rungger and George H. Booth and Jonathan Tennyson}
}

@misc{farhi20A14quantumapproximateoptimizationalgorithm,
  title         = {A Quantum Approximate Optimization Algorithm},
  author        = {Edward Farhi and Jeffrey Goldstone and Sam Gutmann},
  year          = {2014},
  eprint        = {1411.4028},
  archiveprefix = {arXiv},
  primaryclass  = {quant-ph},
  url           = {https://arxiv.org/abs/1411.4028}
}

@misc{chowdhury2016quantumalgorithmsgibbssampling,
  title         = {Quantum algorithms for Gibbs sampling and hitting-time estimation},
  author        = {Anirban Narayan Chowdhury and Rolando D. Somma},
  year          = {2016},
  eprint        = {1603.02940},
  archiveprefix = {arXiv},
  primaryclass  = {quant-ph},
  url           = {https://arxiv.org/abs/1603.02940}
}

@article{Motta2020,
  author  = {Motta, Mario
             and Sun, Chong
             and Tan, Adrian T. K.
             and O'Rourke, Matthew J.
             and Ye, Erika
             and Minnich, Austin J.
             and Brand{\~a}o, Fernando G. S. L.
             and Chan, Garnet Kin-Lic},
  title   = {Determining eigenstates and thermal states on a quantum computer using quantum imaginary time evolution},
  journal = {Nature Physics},
  year    = {2020},
  month   = {Feb},
  day     = {01},
  volume  = {16},
  number  = {2},
  pages   = {205-210},
  issn    = {1745-2481},
  doi     = {10.1038/s41567-019-0704-4},
  url     = {https://doi.org/10.1038/s41567-019-0704-4}
}

@article{SCHOLLWOCK201196,
  title   = {The density-matrix renormalization group in the age of matrix product states},
  journal = {Annals of Physics},
  volume  = {326},
  number  = {1},
  pages   = {96-192},
  year    = {2011},
  note    = {January 2011 Special Issue},
  issn    = {0003-4916},
  doi     = {https://doi.org/10.1016/j.aop.2010.09.012},
  url     = {https://www.sciencedirect.com/science/article/pii/S0003491610001752},
  author  = {Ulrich Schollwöck}
}

@article{PhysRevB.105.075131,
  title     = {Dissipation-assisted operator evolution method for capturing hydrodynamic transport},
  author    = {Rakovszky, Tibor and von Keyserlingk, C. W. and Pollmann, Frank},
  journal   = {Phys. Rev. B},
  volume    = {105},
  issue     = {7},
  pages     = {075131},
  numpages  = {10},
  year      = {2022},
  month     = {Feb},
  publisher = {American Physical Society},
  doi       = {10.1103/PhysRevB.105.075131},
  url       = {https://link.aps.org/doi/10.1103/PhysRevB.105.075131}
}

@inbook{doi:https://doi.org/10.1002/3527603794.ch7,
  author    = {Weigt, Martin},
  publisher = {John Wiley \& Sons, Ltd},
  isbn      = {9783527603794},
  title     = {The Random 3-Satisfiability Problem: From the Phase Transition to the Efficient Generation of Hard, but Satisfiable Problem Instances},
  booktitle = {New Optimization Algorithms in Physics},
  chapter   = {7},
  pages     = {119-137},
  doi       = {https://doi.org/10.1002/3527603794.ch7},
  year      = {2004}
}

@article{PhysRevLett.100.030504,
  title     = {Entropy Scaling and Simulability by Matrix Product States},
  author    = {Schuch, Norbert and Wolf, Michael M. and Verstraete, Frank and Cirac, J. Ignacio},
  journal   = {Phys. Rev. Lett.},
  volume    = {100},
  issue     = {3},
  pages     = {030504},
  numpages  = {4},
  year      = {2008},
  month     = {Jan},
  publisher = {American Physical Society},
  doi       = {10.1103/PhysRevLett.100.030504},
  url       = {https://link.aps.org/doi/10.1103/PhysRevLett.100.030504}
}

@inproceedings{10.1145/237814.237866,
  author    = {Grover, Lov K.},
  title     = {A fast quantum mechanical algorithm for database search},
  year      = {1996},
  isbn      = {0897917855},
  publisher = {Association for Computing Machinery},
  address   = {New York, NY, USA},
  url       = {https://doi.org/10.1145/237814.237866},
  doi       = {10.1145/237814.237866},
  booktitle = {Proceedings of the Twenty-Eighth Annual ACM Symposium on Theory of Computing},
  pages     = {212–219},
  numpages  = {8},
  location  = {Philadelphia, Pennsylvania, USA},
  series    = {STOC '96}
}

@misc{zhang2025optimalschedulemultichannelquantum,
  title         = {Optimal schedule of multi-channel quantum Zeno dragging with application to solving the k-SAT problem},
  author        = {Yipei Zhang and Alain Sarlette and Philippe Lewalle and Tathagata Karmakar and K. Birgitta Whaley},
  year          = {2025},
  eprint        = {2507.16128},
  archiveprefix = {arXiv},
  primaryclass  = {quant-ph},
  url           = {https://arxiv.org/abs/2507.16128}
}

@article{PhysRevA.70.052328,
  title     = {Improved simulation of stabilizer circuits},
  author    = {Aaronson, Scott and Gottesman, Daniel},
  journal   = {Phys. Rev. A},
  volume    = {70},
  issue     = {5},
  pages     = {052328},
  numpages  = {14},
  year      = {2004},
  month     = {Nov},
  publisher = {American Physical Society},
  doi       = {10.1103/PhysRevA.70.052328},
  url       = {https://link.aps.org/doi/10.1103/PhysRevA.70.052328}
}

@article{Obst_2024,
  title     = {Wigner’s theorem for stabilizer states and quantum designs},
  volume    = {65},
  issn      = {1089-7658},
  url       = {http://dx.doi.org/10.1063/5.0222546},
  doi       = {10.1063/5.0222546},
  number    = {11},
  journal   = {Journal of Mathematical Physics},
  publisher = {AIP Publishing},
  author    = {Obst, Valentin and Heimendahl, Arne and Singal, Tanmay and Gross, David},
  year      = {2024},
  month     = nov
}

@article{PhysRevA.71.022316,
  title     = {Universal quantum computation with ideal Clifford gates and noisy ancillas},
  author    = {Bravyi, Sergey and Kitaev, Alexei},
  journal   = {Phys. Rev. A},
  volume    = {71},
  issue     = {2},
  pages     = {022316},
  numpages  = {14},
  year      = {2005},
  month     = {Feb},
  publisher = {American Physical Society},
  doi       = {10.1103/PhysRevA.71.022316},
  url       = {https://link.aps.org/doi/10.1103/PhysRevA.71.022316}
}

@article{Leone_2024,
  title     = {Stabilizer entropies are monotones for magic-state resource theory},
  volume    = {110},
  issn      = {2469-9934},
  url       = {http://dx.doi.org/10.1103/PhysRevA.110.L040403},
  doi       = {10.1103/physreva.110.l040403},
  number    = {4},
  journal   = {Physical Review A},
  publisher = {American Physical Society (APS)},
  author    = {Leone, Lorenzo and Bittel, Lennart},
  year      = {2024},
  month     = oct
}

@article{PhysRevB.110.045101,
  title     = {Nonstabilizerness versus entanglement in matrix product states},
  author    = {Frau, M. and Tarabunga, P. S. and Collura, M. and Dalmonte, M. and Tirrito, E.},
  journal   = {Phys. Rev. B},
  volume    = {110},
  issue     = {4},
  pages     = {045101},
  numpages  = {13},
  year      = {2024},
  month     = {Jul},
  publisher = {American Physical Society},
  doi       = {10.1103/PhysRevB.110.045101},
  url       = {https://link.aps.org/doi/10.1103/PhysRevB.110.045101}
}

@article{PhysRevLett.128.050402,
  title     = {Stabilizer R\'enyi Entropy},
  author    = {Leone, Lorenzo and Oliviero, Salvatore F. E. and Hamma, Alioscia},
  journal   = {Phys. Rev. Lett.},
  volume    = {128},
  issue     = {5},
  pages     = {050402},
  numpages  = {5},
  year      = {2022},
  month     = {Feb},
  publisher = {American Physical Society},
  doi       = {10.1103/PhysRevLett.128.050402},
  url       = {https://link.aps.org/doi/10.1103/PhysRevLett.128.050402}
}

@article{Haug2023stabilizerentropies,
  doi       = {10.22331/q-2023-08-28-1092},
  url       = {https://doi.org/10.22331/q-2023-08-28-1092},
  title     = {Stabilizer entropies and nonstabilizerness monotones},
  author    = {Haug, Tobias and Piroli, Lorenzo},
  journal   = {{Quantum}},
  issn      = {2521-327X},
  publisher = {{Verein zur F{\"{o}}rderung des Open Access Publizierens in den Quantenwissenschaften}},
  volume    = {7},
  pages     = {1092},
  month     = aug,
  year      = {2023}
}

@article{PhysRevLett.131.180401,
  title     = {Nonstabilizerness via Perfect Pauli Sampling of Matrix Product States},
  author    = {Lami, Guglielmo and Collura, Mario},
  journal   = {Phys. Rev. Lett.},
  volume    = {131},
  issue     = {18},
  pages     = {180401},
  numpages  = {6},
  year      = {2023},
  month     = {Oct},
  publisher = {American Physical Society},
  doi       = {10.1103/PhysRevLett.131.180401},
  url       = {https://link.aps.org/doi/10.1103/PhysRevLett.131.180401}
}

@article{PRXQuantum.4.010301,
  title     = {Scalable Measures of Magic Resource for Quantum Computers},
  author    = {Haug, Tobias and Kim, M.S.},
  journal   = {PRX Quantum},
  volume    = {4},
  issue     = {1},
  pages     = {010301},
  numpages  = {23},
  year      = {2023},
  month     = {Jan},
  publisher = {American Physical Society},
  doi       = {10.1103/PRXQuantum.4.010301},
  url       = {https://link.aps.org/doi/10.1103/PRXQuantum.4.010301}
}

@misc{gottesman1998heisenbergrepresentationquantumcomputers,
  title         = {The Heisenberg Representation of Quantum Computers},
  author        = {Daniel Gottesman},
  year          = {1998},
  eprint        = {quant-ph/9807006},
  archiveprefix = {arXiv},
  primaryclass  = {quant-ph},
  url           = {https://arxiv.org/abs/quant-ph/9807006}
}

@article{PRXQuantum.4.040317,
  title     = {Many-Body Magic Via Pauli-Markov Chains---From Criticality to Gauge Theories},
  author    = {Tarabunga, Poetri Sonya and Tirrito, Emanuele and Chanda, Titas and Dalmonte, Marcello},
  journal   = {PRX Quantum},
  volume    = {4},
  issue     = {4},
  pages     = {040317},
  numpages  = {19},
  year      = {2023},
  month     = {Oct},
  publisher = {American Physical Society},
  doi       = {10.1103/PRXQuantum.4.040317},
  url       = {https://link.aps.org/doi/10.1103/PRXQuantum.4.040317}
}

@article{PhysRevLett.88.188701,
  title     = {Hiding Solutions in Random Satisfiability Problems: A Statistical Mechanics Approach},
  author    = {Barthel, W. and Hartmann, A. K. and Leone, M. and Ricci-Tersenghi, F. and Weigt, M. and Zecchina, R.},
  journal   = {Phys. Rev. Lett.},
  volume    = {88},
  issue     = {18},
  pages     = {188701},
  numpages  = {4},
  year      = {2002},
  month     = {Apr},
  publisher = {American Physical Society},
  doi       = {10.1103/PhysRevLett.88.188701},
  url       = {https://link.aps.org/doi/10.1103/PhysRevLett.88.188701}
}

@misc{szombathy2025independentstabilizerrenyientropy,
  title         = {Independent stabilizer R\'enyi entropy and entanglement fluctuations in random unitary circuits},
  author        = {Dominik Szombathy and Angelo Valli and Cătălin Paşcu Moca and Lóránt Farkas and Gergely Zaránd},
  year          = {2025},
  eprint        = {2501.11489},
  archiveprefix = {arXiv},
  primaryclass  = {quant-ph},
  url           = {https://arxiv.org/abs/2501.11489}
}

@article{doi:10.3233/SAT190039,
  author  = {Armin Biere},
  title   = {PicoSAT Essentials},
  journal = {Journal on Satisfiability, Boolean Modelling and Computation},
  volume  = {4},
  number  = {2-4},
  pages   = {75-97},
  year    = {2008},
  doi     = {10.3233/SAT190039},
  url     = { 
             
             
             https://journals.sagepub.com/doi/abs/10.3233/SAT190039
             
             
             }
}

@article{10.21468/SciPostPhysCodeb.4,
  title     = {{The ITensor Software Library for Tensor Network Calculations}},
  author    = {Matthew Fishman and Steven R. White and E. Miles Stoudenmire},
  journal   = {SciPost Phys. Codebases},
  pages     = {4},
  year      = {2022},
  publisher = {SciPost},
  doi       = {10.21468/SciPostPhysCodeb.4},
  url       = {https://scipost.org/10.21468/SciPostPhysCodeb.4}
}

@article{10.21468/SciPostPhysCodeb.4-r0.3,
  title     = {{Codebase release 0.3 for ITensor}},
  author    = {Matthew Fishman and Steven R. White and E. Miles Stoudenmire},
  journal   = {SciPost Phys. Codebases},
  pages     = {4-r0.3},
  year      = {2022},
  publisher = {SciPost},
  doi       = {10.21468/SciPostPhysCodeb.4-r0.3},
  url       = {https://scipost.org/10.21468/SciPostPhysCodeb.4-r0.3}
}

@article{haegeman_2025_17088820,
  author    = {Haegeman, Jutho},
  title     = {KrylovKit},
  month     = sep,
  year      = 2025,
  publisher = {Zenodo},
  journal   = {Zenodo},
  version   = {v0.10.1},
  doi       = {10.5281/zenodo.17088820},
  url       = {https://doi.org/10.5281/zenodo.17088820}
}

@article{mogensen2018optim,
  author  = {Mogensen, Patrick Kofod and Riseth, Asbj{\o}rn Nilsen},
  title   = {Optim: A mathematical optimization package for {Julia}},
  journal = {Journal of Open Source Software},
  year    = {2018},
  volume  = {3},
  number  = {24},
  pages   = {615},
  doi     = {10.21105/joss.00615}
}

@article{McArdle2019,
  author   = {McArdle, Sam
              and Jones, Tyson
              and Endo, Suguru
              and Li, Ying
              and Benjamin, Simon C.
              and Yuan, Xiao},
  title    = {Variational ansatz-based quantum simulation of imaginary time evolution},
  journal  = {npj Quantum Information},
  year     = {2019},
  month    = {Sep},
  day      = {06},
  volume   = {5},
  number   = {1},
  pages    = {75},
  issn     = {2056-6387},
  doi      = {10.1038/s41534-019-0187-2},
  url      = {https://doi.org/10.1038/s41534-019-0187-2}
}

@misc{anglescastillo2025understandingquantumimaginarytime,
  title         = {Understanding Quantum Imaginary Time Evolution and its Variational form},
  author        = {Andreu Anglés-Castillo and Luca Ion and Tanmoy Pandit and Rafael Gomez-Lurbe and Rodrigo Martínez and Miguel Angel Garcia-March},
  year          = {2025},
  eprint        = {2510.02015},
  archiveprefix = {arXiv},
  primaryclass  = {quant-ph},
  url           = {https://arxiv.org/abs/2510.02015}
}

@article{PhysRevB.110.245109,
  title     = {Geometrically taming dynamical entanglement growth in purified quantum states},
  author    = {Pokart, Tim and Lehmann, Carl and Budich, Jan Carl},
  journal   = {Phys. Rev. B},
  volume    = {110},
  issue     = {24},
  pages     = {245109},
  numpages  = {17},
  year      = {2024},
  month     = {Dec},
  publisher = {American Physical Society},
  doi       = {10.1103/PhysRevB.110.245109},
  url       = {https://link.aps.org/doi/10.1103/PhysRevB.110.245109}
}

@article{tenpy2024,
  title     = {{Tensor network Python (TeNPy) version 1}},
  author    = {Johannes Hauschild and Jakob Unfried and Sajant Anand and Bartholomew Andrews and Marcus Bintz and Umberto Borla and Stefan Divic and Markus Drescher and Jan Geiger and Martin Hefel and Kévin Hémery and Wilhelm Kadow and Jack Kemp and Nico Kirchner and Vincent S. Liu and Gunnar Möller and Daniel Parker and Michael Rader and Anton Romen and Samuel Scalet and Leon Schoonderwoerd and Maximilian Schulz and Tomohiro Soejima and Philipp Thoma and Yantao Wu and Philip Zechmann and Ludwig Zweng and Roger S. K. Mong and Michael P. Zaletel and Frank Pollmann},
  journal   = {SciPost Phys. Codebases},
  pages     = {41},
  year      = {2024},
  publisher = {SciPost},
  doi       = {10.21468/SciPostPhysCodeb.41},
  url       = {https://scipost.org/10.21468/SciPostPhysCodeb.41}
}

@article{PhysRevLett.71.1291,
  title     = {Average entropy of a subsystem},
  author    = {Page, Don N.},
  journal   = {Phys. Rev. Lett.},
  volume    = {71},
  issue     = {9},
  pages     = {1291--1294},
  numpages  = {0},
  year      = {1993},
  month     = {Aug},
  publisher = {American Physical Society},
  doi       = {10.1103/PhysRevLett.71.1291},
  url       = {https://link.aps.org/doi/10.1103/PhysRevLett.71.1291}
}

@book{johnson1979computers,
  title={Computers and intractability: A guide to the theory of NP-completeness},
  author={Johnson, David S and Garey, Michael R},
  year={1979},
  publisher={WH Freeman},
chapter = {6.3},
    isbn={0-7167-1044-7},
  nolink={}
}

@BOOK{Cormen2022-xl,
  title     = "Introduction to Algorithms, fourth edition",
  author    = "Cormen, Thomas H and Leiserson, Charles E",
  publisher = "MIT Press",
  month     =  apr,
  year      =  2022,
  address   = "London, England",
  chapter = {35},
  isbn={ 978-0262046305},
  nolink={}
}

@book{donald2015art,
  title={The Art of Computer Programming: Combinatorial Algorithms},
volume={4B},
  author={Donald E. Knuth},
  year={2022},
  publisher={Addison-Wesley},
  isbn={978-0-201-03806-4},
  chapter={7.2.2.2},
  nolink={}
}

@Article{Fannes1992,
author={Fannes, M.
and Nachtergaele, B.
and Werner, R. F.},
title={Finitely correlated states on quantum spin chains},
journal={Communications in Mathematical Physics},
year={1992},
month={Mar},
day={01},
volume={144},
number={3},
pages={443-490},
issn={1432-0916},
doi={10.1007/BF02099178},
url={https://doi.org/10.1007/BF02099178}
}

@article{doi:10.1137/S0097539704445226,
author = {Kempe, Julia and Kitaev, Alexei and Regev, Oded},
title = {The Complexity of the Local Hamiltonian Problem},
journal = {SIAM Journal on Computing},
volume = {35},
number = {5},
pages = {1070-1097},
year = {2006},
doi = {10.1137/S0097539704445226},

URL = { 
    
        https://doi.org/10.1137/S0097539704445226
    
    

},
eprint = { 
        https://doi.org/10.1137/S0097539704445226
}
}

@article{Aharonov2009,
author={Aharonov, Dorit
and Gottesman, Daniel
and Irani, Sandy
and Kempe, Julia},
title={The Power of Quantum Systems on a Line},
journal={Communications in Mathematical Physics},
year={2009},
month={Apr},
day={01},
volume={287},
number={1},
pages={41-65},
issn={1432-0916},
doi={10.1007/s00220-008-0710-3},
url={https://doi.org/10.1007/s00220-008-0710-3}
}

@article{PhysRevApplied.19.034052,
  title = {Lower Bound for the $T$ Count Via Unitary Stabilizer Nullity},
  author = {Jiang, Jiaqing and Wang, Xin},
  journal = {Phys. Rev. Appl.},
  volume = {19},
  issue = {3},
  pages = {034052},
  numpages = {19},
  year = {2023},
  month = {Mar},
  publisher = {American Physical Society},
  doi = {10.1103/PhysRevApplied.19.034052},
  url = {https://link.aps.org/doi/10.1103/PhysRevApplied.19.034052}
}

@article{Ruiz2025,
author={Ruiz, Francisco J. R.
and Laakkonen, Tuomas
and Bausch, Johannes
and Balog, Matej
and Barekatain, Mohammadamin
and Heras, Francisco J. H.
and Novikov, Alexander
and Fitzpatrick, Nathan
and Romera-Paredes, Bernardino
and van de Wetering, John
and Fawzi, Alhussein
and Meichanetzidis, Konstantinos
and Kohli, Pushmeet},
title={Quantum circuit optimization with AlphaTensor},
journal={Nature Machine Intelligence},
year={2025},
month={Mar},
day={01},
volume={7},
number={3},
pages={374-385},
issn={2522-5839},
doi={10.1038/s42256-025-01001-1},
url={https://doi.org/10.1038/s42256-025-01001-1}
}

@misc{gidney2024magicstatecultivationgrowing,
      title={Magic state cultivation: growing T states as cheap as CNOT gates}, 
      author={Craig Gidney and Noah Shutty and Cody Jones},
      year={2024},
      eprint={2409.17595},
      archivePrefix={arXiv},
      primaryClass={quant-ph},
      url={https://arxiv.org/abs/2409.17595}, 
}

@article{pokart_2026_18743084,
  author       = {Pokart, Tim and
                  Pollmann, Frank and
                  Budich, Jan Carl},
  title        = {Entanglement Barriers from Computational
                   Complexity: Matrix-Product-State Approach to
                   Satisfiability
                  },
  month        = feb,
  year         = 2026,
  publisher    = {Zenodo},
  journal      = {Zenodo},
  doi          = {10.5281/zenodo.18743084},
  url          = {https://doi.org/10.5281/zenodo.18743084},
}

@misc{preisser2025variationalmatrixproductstates,
      title={Variational (matrix) product states for combinatorial optimization}, 
      author={Guillermo Preisser and Conor Mc Keever and Michael Lubasch},
      year={2025},
      eprint={2512.20613},
      archivePrefix={arXiv},
      primaryClass={quant-ph},
      url={https://arxiv.org/abs/2512.20613}, 
}

\end{document}